\documentclass[superscriptaddress,twocolumn,aps,showpacs,pra,preprintnumbers]{revtex4-1}
\usepackage{epsfig}
\usepackage{graphicx}
\usepackage{amsmath}
\usepackage{bbm}
\usepackage{setspace}
\usepackage[colorlinks=true,citecolor=blue,linkcolor=blue]{hyperref}

\usepackage[usenames,dvipsnames]{xcolor}     
\usepackage{soul}       

\newcommand{\adrfont}{\fontfamily{ptm}\selectfont}
\newcommand{\LENS}{\adrfont LENS and Dipartimento di Fisica e Astronomia, Universit\'a di Firenze, 50019 Sesto Fiorentino, Italy}
\newcommand{\INO} {\adrfont Istituto Nazionale di Ottica, CNR, 50019 Sesto Fiorentino, Italy}
\newcommand{\PSUD} {\adrfont LPTMS, CNRS, Univ. Paris-Sud, Universit\'e Paris-Saclay, 91405 Orsay, France}

\begin{document}

\title{Finite-temperature effects on interacting  bosonic 1D systems in disordered lattices}

\author{Lorenzo Gori}
\affiliation{\LENS}
\author{Thomas Barthel}
\affiliation{\adrfont Department of Physics, Duke University, Durham, North Carolina 27708, USA}
\affiliation{\PSUD}
\author{Avinash Kumar}
\affiliation{\LENS}
\author{Eleonora Lucioni}
\affiliation{\LENS}
\affiliation{\INO}
\author{Luca Tanzi}
\affiliation{\LENS}
\author{Massimo Inguscio}
\affiliation{\LENS}
\affiliation{\adrfont Istituto Nazionale di Ricerca Metrologica, 10135 Torino, Italy}
\author{Giovanni Modugno}
\affiliation{\LENS}
\affiliation{\INO}
\author{Thierry Giamarchi}
\affiliation{\adrfont Department of Quantum Matter Physics, University of Geneva, 1211 Geneva, Switzerland}
\author{Chiara D'Errico}
\affiliation{\LENS}
\affiliation{\INO}
\author{Guillaume Roux}
\affiliation{\PSUD}
\date{\today}

\begin{abstract}
We analyze the finite-temperature effects on the phase diagram describing the insulating properties of interacting 1D bosons in a quasi-periodic lattice. We examine thermal effects by comparing experimental results to exact diagonalization for small-sized systems and to density-matrix renormalization group (DMRG) computations.
At weak interactions, we find short thermal correlation lengths, indicating a substantial impact of temperature on the system coherence. Conversely, at strong interactions, the obtained thermal correlation lengths are significantly larger than the localization length, and the quantum nature of the $T=0$ Bose glass phase is preserved up to a crossover temperature that depends on the disorder strength.
Furthermore, in the absence of disorder, we show how quasi-exact finite-$T$ DMRG computations, compared to experimental results, can be employed to estimate the temperature, which is not directly accessible in the experiment. 
\end{abstract}


\maketitle


\section{Introduction}

For their ability to simulate condensed matter systems, ultracold  atoms in disordered optical potentials are known to be very effective and versatile systems. The appeal of such systems, already highlighted in the observation of Anderson localization \cite{Anderson, Roati08, Billy} for vanishing interactions, is increasing in the research activity on many-body quantum physics. Since several decades, large effort has been made to investigate the combined effect of disorder and interaction on the insulating properties of one-dimensional (1D) bosonic systems, both theoretically and experimentally.

From a theoretical view point, the $T=0$ phase diagram describing the superfluid-insulator transitions has been studied for both random disorder \cite{Giamarchi88,Prokofev98,Rapsch99} and quasi-periodic lattices \cite{Roth,Roscilde,Deng,Roux, Modugno}. The quasi-periodic lattice displays behaviors that are qualitatively and quantitatively different from those of a true random disorder. Yet, the occurrence of localization makes it a remarkable testbed for studying the Bose-glass physics.
On the experimental side, the disorder-interaction phase diagram has been examined \cite{Fallani, Deissler, DeMarco} and, in the recent study of Ref.~\cite{Derrico}, measurements of momentum distribution, transport and excitation spectra showed a finite-$T$ reentrant insulator resembling the one predicted by theory.

In this context, the question of the effect of finite temperature is however still open \cite{Michal} and a direct link between the $T=0$ theory and the experiment is still missing. In particular, whether and to what extent the $T=0$ quantum phases persist at the low but finite experimental temperatures still has to be understood. 
Increasing the temperature in a clean (i.e., non-disordered) system, the quantum Mott domains progressively shrink, vanishing at the ``melting'' temperature $k_B T \simeq 0.2U$, with $U$ being the Mott energy gap \cite{Gerbier}. 
In the presence of disorder, no theoretical predictions are so far available.

In this article, starting from the recent experimental study \cite{Derrico}, we analyze  the coherence properties of the system. By comparing the experimental finite-$T$ data with a phenomenological approach based on DMRG calculations \cite{White1992,White1993,Schollwock} for our inhomogeneous system at $T=0$, we provide a qualitative estimation of the coherence loss induced by temperature throughout the disorder-interaction diagram. In this framework, the coherence loss is quantified in terms of a phenomenological parameter, the effective thermal correlation length. Furthermore, a  rigorous analysis of the temperature dependence of the correlation length  is provided by exact diagonalization of the Hamiltonian for the case of small homogeneous systems. A reduction of the correlation length above a disorder-dependent characteristic temperature can be interpreted as a crossover from a quantum to a normal phase.  
 In the regime of strong interactions, the exact diagonalization method -- which well reproduces the melting temperature for the clean commensurate Mott insulator -- is found to  apply also to the disordered case, thus providing a crossover temperature for  the incommensurate Bose glass phase.  

Complementarily, we show how to estimate the temperature of the experimental system by comparison of the measured momentum distribution with quasi-exact theoretical results, obtained with a finite-$T$ DMRG method \cite{Verstraete,Feiguin,Barthel,Binder}. Up to now it was possible to determine the temperature of a 1D quasi-condensate in the presence of the trap alone \cite{Gerbier03}. 
By using the DMRG simulations, it is also possible to determine temperatures of quasi-1D systems in the presence of lattice potentials. For the present experiment we estimate the temperature in the superfluid regime
without disorder.
Problems can arise in the analysis  of insulating experimental systems as these are not necessarily in thermal equilibrium. Attempts of temperature measurements for such systems are reported as well, highlighting the difficulties also caused  by the coexistence of different phases in the considered inhomogeneous system.    

The exposition of this work is organized as follows. 
Sec.~\ref{sec:exp} describes the experimental setup and methods. 
In Sec.~\ref{sec:methods}, we explain the theoretical methods employed in the subsequent sections to analyze the finite-$T$ effects on the quantum phases of the system.
After recalling the main experimental results reported in Ref.~\cite{Derrico}, Sec.~\ref{sec:phenomenological}  presents a phenomenological approach based on $T=0$ DMRG calculations  that captures thermal effects and introduces an effective thermal correlation length. The effect of the system inhomogeneity is analyzed as well.
In Sec.~\ref{sec:ED}, we perform exact diagonalization for small homogeneous systems. For weak  interactions, this provides the $T$-dependence of the correlation length for the superfluid and weakly interacting Bose glass while, for strong interactions, it provides the crossover temperature for the existence of the quantum phases, the Mott insulator and the strongly interacting Bose glass. Measurements of the system entropy support the latter results. 
In Sec.~\ref{sec:thermometry}, we use finite-$T$ DMRG calculations for an \emph{ab initio} thermometry in a clean system. In particular, experimental temperatures are estimated by comparing  the experimental momentum distributions with quasi-exact DMRG calculations. 
In Sec.~\ref{sec:experimentaltemperatures}, entropy measurements throughout the full disorder-interaction diagram  are also provided. Finally, the conclusions are reported in Sec.~\ref{sec:conclusions}.

\section{Experimental methods}
\label{sec:exp}
\begin{figure}[t]
\centering
\includegraphics[width=0.9\columnwidth] {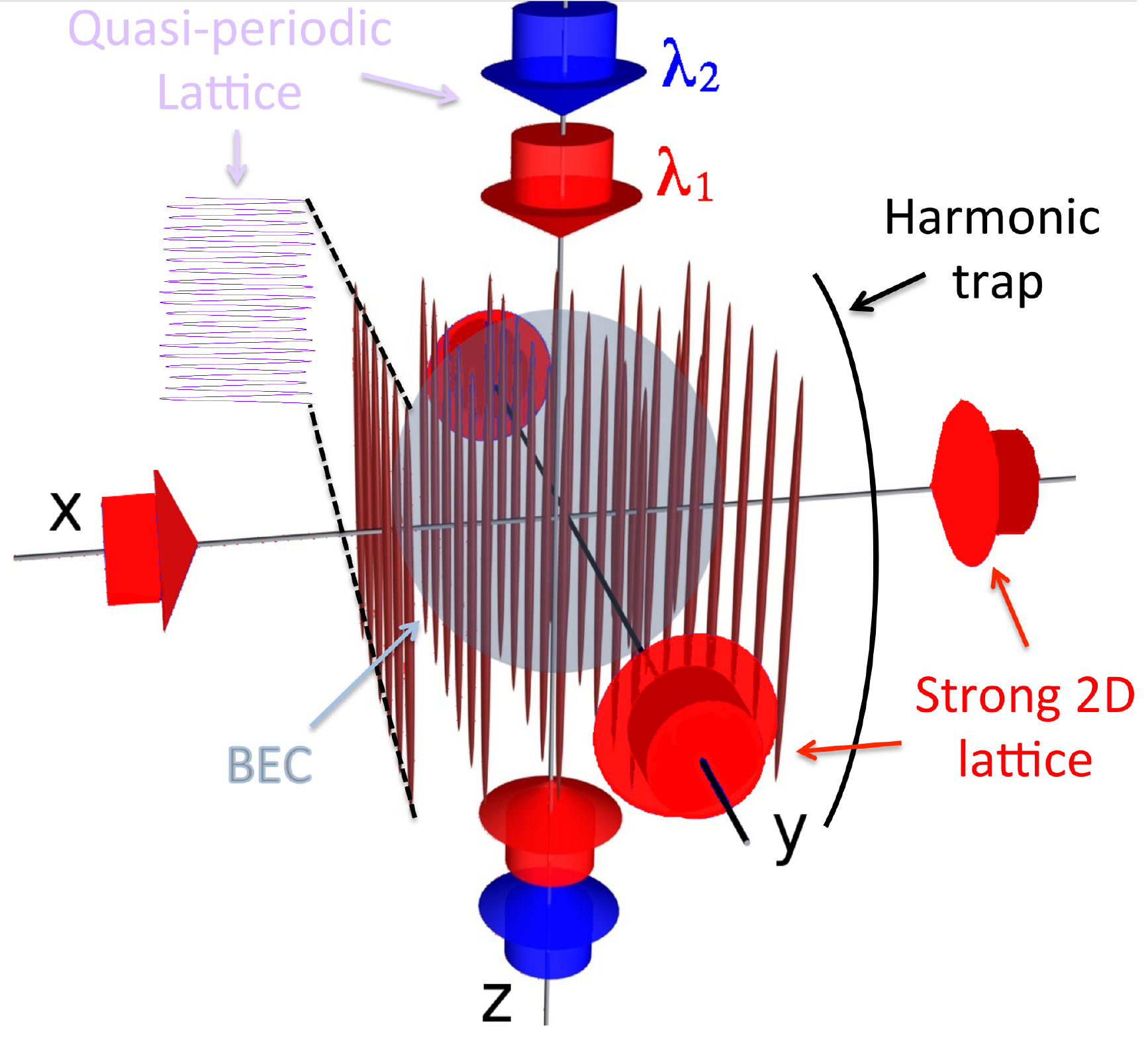}
\caption{Experimental setup. Two horizontal optical lattices provide a tight confinement forming an array of 1D vertical potential tubes for the $^{39}$K atoms with tunable interaction energy $U$. The vertical quasi-periodic  potential is formed by superimposing two incommensurate optical lattices: the main lattice ($\lambda_1 = 1064$\;nm), which is related to the tunneling energy $J$, and the secondary one ($\lambda_2 =  859.6$\;nm), which is related to the disorder amplitude $\Delta$. The harmonic trapping confinement makes the 1D systems inhomogeneous.}
\label{setup}
\end{figure}

Starting from a 3D Bose-Einstein condensate (BEC) with $N_{\text{tot}}\simeq 35\,000$ atoms of $^{39}$K, a strong horizontal 2D optical lattice (with depth of 30 recoil energies) is ramped up such that an array of independent potential tubes directed along the $z$-axis is created. This forms a set of about 500 quasi-1D systems, as depicted in Fig.~\ref{setup}.
Additionally, a quasi-periodic lattice along the $z$-direction is then ramped up, yielding a set of  disordered quasi-1D systems \cite{Fallani,Roati08}. 
Such systems are described by the disordered Bose-Hubbard Hamiltonian  \cite{Roux,Roscilde}
\begin{multline}\label{eq1}
	H = -J \sum_i(b_i^{\dag}b^{\phantom{\dag}}_{i+1}+h.c.) + \Delta \sum_i \cos(2 \pi \delta i)n_i\\
	+ \frac{U}{2} \sum_i n_i (n_i-1) + \frac{\alpha}{2} \sum_i(i-i_0)^2 n_i,
\end{multline}
where $b_i^{\dag}$, $b_i$, and $n_i=b_i^{\dag}b^{\phantom{\dag}}_i$ are  the creation, annihilation and number operators at site $i$.
The Hamiltonian is characterized by three main energy scales: the tunneling energy $J$, the quasi-disorder strength $\Delta$ and the interaction energy $U$. The tunneling rate $J/h \simeq 110\;\text{Hz}$ is set by the depth of the primary lattice with spacing $d=\lambda_1/2 = 0.532\;\mu\text{m}$. $\Delta$ can be suitably varied by changing the depth of a weaker secondary lattice, superimposed to the primary one and having an incommensurate wavelength $\lambda_2$ such that the ratio $\delta=\lambda_1/\lambda_2=1.243\ldots$ is far from a simple fraction and  mimics  the potential that would be created by a truly irrational number. $U$ can be easily controlled as well thanks to a broad Feshbach resonance \cite{Roati07} which allows to change the inter-particle scattering length $a_s$ from about zero to large positive values. 
Finally, the fourth term of the Hamiltonian, which is characterized by the parameter $\alpha\simeq 0.26 J$, represents the harmonic trapping potential, centered around  lattice site $i_0$. Depending on the value of $U$, the mean site occupancy can range from $n=2$ to $n=8$. 
More details on the experimental apparatus and procedures are given in Ref.~\cite{Derrico}.

Theoretical phase diagrams for the model \eqref{eq1} were obtained by numerical computation and analytical arguments \cite{Roth,Roscilde,Deng,Roux} for the ideal case of zero temperature and no trapping potential. 
However, due to experimental constrains, the 1D quasi-condensates we actually produce are at low but finite temperatures (of the order of few $J$, thus below the characteristic degeneracy temperature $T_D\simeq8J/k_B$ \cite{Petrov}). Moreover, the unavoidable trapping confinement used in the experiment makes the system inhomogeneous and limits its size. As a result, in the experimental system, different phases coexist and the theoretical sharp quantum phase transitions occurring in the case of the thermodynamic limit are actually  replaced by broad crossovers.

The analysis of the next sections is mainly based on the momentum distribution $P(k)$.  Experimentally, $P(k)$ is obtained by releasing the atomic cloud from the trapping potential and letting it expand freely  for 16 ms before acquiring an absorption image. From the root-mean-square (rms) width of $P(k)$ we get information about the coherence of the system.

\section{Theoretical methods}
\label{sec:methods}

\subsection{Averaged momentum distribution}
DMRG calculations, as described in subsections \ref{sec:phenomAnsatz} and \ref{sec:methodDMRG-finiteT}, give access to the density profiles in the 1D tubes and to the single-particle correlation functions $g_{ij}(T)=\langle b_i^\dagger b_j\rangle_{T}$,  where $\langle\cdots\rangle_T$ denotes the quantum-mechanical expectation value in thermal equilibrium. The corresponding momentum distributions are computed according to
\begin{equation} \label{eq:MomentumDistribution}
P(k) = |W(k)|^2 \sum_{i,j} e^{ik(i-j)} \bar{g}_{ij}\;,
\end{equation}
where $W(k)$ is the Fourier transform of the numerically computed Wannier function. For quasi-momenta $k$ in the first Brillouin zone, $W(k)$ can be approximated very well by an inverse parabola. The notation $\overline{(\cdots)}$ indicates the average over all tubes in the setup. 

\subsection{Distribution of particles among tubes}
\label{sec:distrib}
There are several assumptions made in modeling the experimental setups. As numerical calculations and most theoretical analyses are better suited for studying lattice models, one has to derive the lattice model from the continuous Hamiltonian corresponding to the optical lattices setup. For our system, this issue is discussed in Refs.~\cite{Roscilde,Roux}.

The experiment comprises a collection of 1D tubes modeled by Hamiltonian \eqref{eq1}. Due to the transverse component of the harmonic trapping potential, these tubes contain different numbers of particles. The total number of particles $N_{\text{tot}}$ is  known with an uncertainty of 15\% and the distribution of particles among the tubes is also not exactly known. In the theoretical analysis, we consider two different distributions, that we call Thomas-Fermi (TF) distribution and grand-canonical (GC) distribution, respectively. The former basically assumes that, during the ramping of the lattice potentials, particles are not redistributed among the tubes. The latter rather assumes that the system evolves until it has reached its equilibrium state and particles have correspondingly
redistributed between the tubes.

For the Thomas-Fermi approximation, the distribution of particles among the tubes still corresponds to the Thomas-Fermi distribution of  the anisotropic 3D BEC  before the ramping of the lattice potentials. Integrating the Thomas-Fermi profile along the $z$-direction gives a continuous 2D transverse density profile of the form
\begin{equation}
	\label{eq:TFdistribution}
	N(\mathbf{r}_\bot) = N_{\text{max}}\left(1-\frac{\mathbf{r}_\bot^2}{R_r^2}\right)^{3/2},
\end{equation}
where $R_r = \sqrt{{2\mu}/{m\omega_r^2}}$ and $\mu = \frac{\hbar\bar{\omega}}{2} \left(15{a_s}/{\bar{a}}\right)^{2/5} N_{\text{tot}}^{2/5}$. 
Here $\omega_r$ and $\bar{\omega}$ are the radial and mean optical trap frequencies before the loading of the tubes, and  $\bar{a}=\sqrt{\hbar/m \bar{\omega}}$ is the associated harmonic length. Inserting the experimental parameters, we obtain the relation $R_r \simeq 1.9\,N_{\text{tot}}^{2/5} d$. The number of atoms in the central tube is given by $N_{\text{max}} \simeq \frac{5}{2\pi}\frac{d^2}{R_r^2} N_{\text{tot}}$. For DMRG computations, we approximate Eq.~\eqref{eq:TFdistribution} by a set of integers $\{N_\nu\}$ satisfying $\sum_\nu N_\nu=N_{\text{tot}}$, where $N_\nu$ denotes the number of particles in tube $\nu$. In this approach, the distribution of particles  depends only on $N_{\text{tot}}$ and not on $T$, $U$, $J$, or $\Delta$.

In addition, we consider the grand-canonical approach, which is well suited for calculations done with finite-$T$ DMRG. This is also useful in the classical limit ($J=0$) for which the grand partition function naturally factorizes. We choose a global chemical potential $\mu$ such that the expectation value of the total number of particles is $N_{\text{tot}}$. As the different tubes are independent of each other, the effective chemical potential $\mu_{\nu}$  of tube $\nu$ is determined by $\mu$ and by the transverse component of the harmonic trapping potential such that $\mu_\nu = \mu - \frac12 m\omega_r^2 \mathbf{r}_{\bot,\nu}^2$ where $\mathbf{r}_{\bot,\nu}$ is the transverse 2D position of tube $\nu$. Physically, this assumes that particles are redistributed between tubes when the lattice potentials are ramped up. In order to determine $\mu$ for a given total number of particles $N_{\text{tot}}=\sum_\nu N(\mu_\nu)$, we rely on data for the number of atoms $N(\mu_\nu)$ in a tube for a given chemical potential of the tube. $N(\mu_\nu)$ is computed numerically with finite-$T$ DMRG or in the classical limit of the model. Contrary to the TF approach, $N(\mu_\nu)$ here depends on the temperature and on all parameters of the model, in particular the interaction. As in the experiment, theoretical expectation values are averaged over all tubes.
\begin{figure}[h]
\centering
\includegraphics[width=\columnwidth] {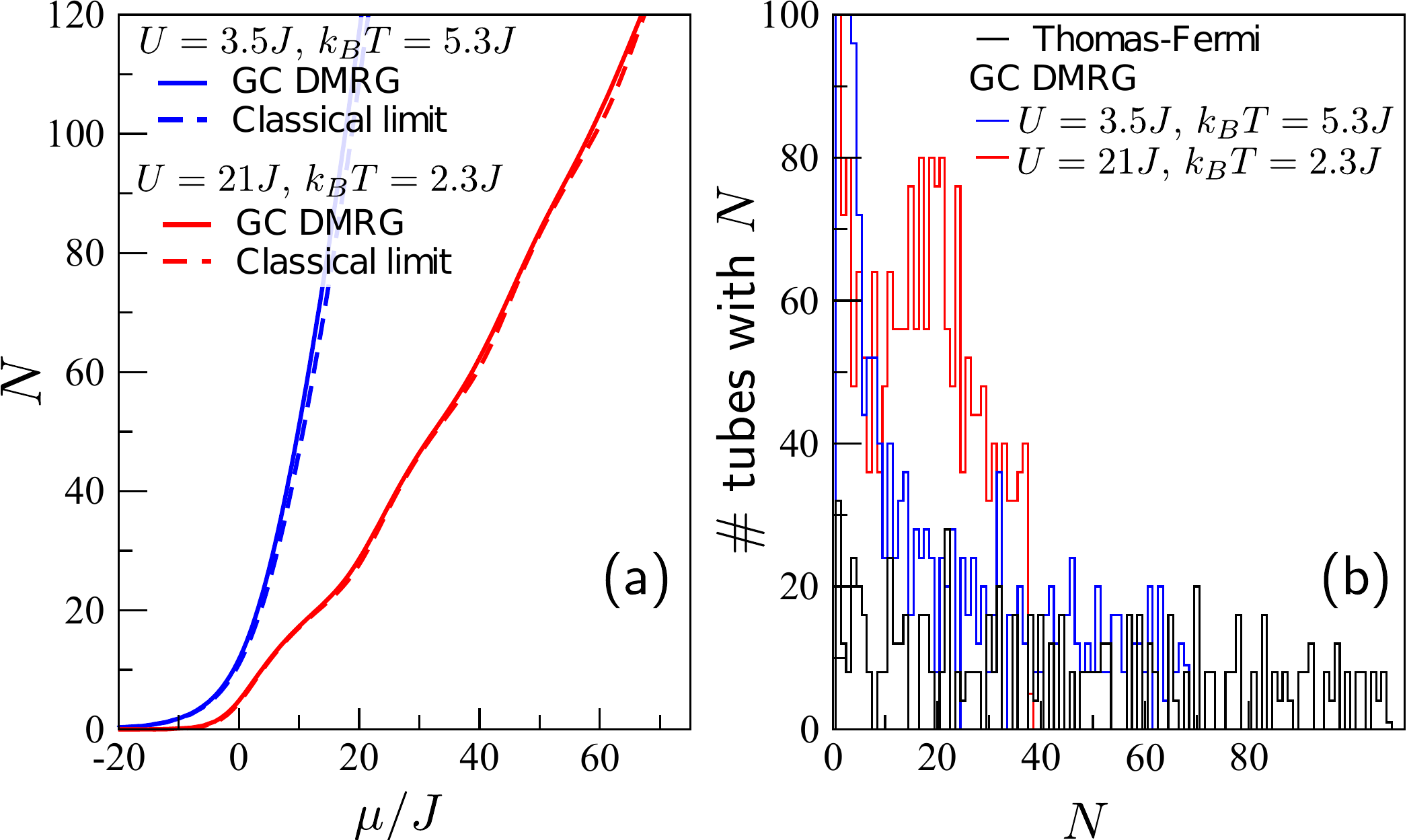}
\caption{\textsf{(a)} Finite-$T$ DMRG data and classical limit for the number of atoms $N(\mu)$ in a tube, as a function of the tube chemical potential $\mu$. \textsf{(b)} Distribution of the tube particle numbers. Compared to the grand-canonical distribution of particles among tubes, the Thomas-Fermi distribution favors more highly-filled tubes in the center of the trap.}
\label{fig:NofMu}
\end{figure}

Typical $N(\mu_\nu)$ relations for the trapped system are shown in Fig.~\ref{fig:NofMu}a for the values of interaction and temperature that will be used later. The corresponding distribution of the atom numbers in the tubes is given in Fig.~\ref{fig:NofMu}b, showing that, in comparison to the TF approximation, the GC approach favors tubes with lower fillings. 
For the typical parameters of the experiment and range of temperatures found hereafter, the modification of  $P(k)$ due to a change of $N_{\text{tot}}$ by $\pm 15\%$ is less relevant than the modification obtained by changing the assumption about the tube atom number distribution (TF or GC). Consequently, unless stated differently, we use $N_{\text{tot}} = 35\;000$ in the following.

\subsection{Phenomenological finite-\texorpdfstring{$T$}{T} approach based on \texorpdfstring{$T=0$}{T=0} DMRG}
\label{sec:phenomAnsatz}
For a single tube, standard DMRG \cite{White1992,White1993,Schollwock} calculations provide accurate $T=0$ results for the momentum distribution.
As the analysis of the full $U$-$\Delta$ diagram requires computations for 94 points, a systematic scan of the temperature for each point using finite-$T$ DMRG represents a numerical challenge. In the case of the 2D Bose-Hubbard model without disorder, such an \emph{ab initio} fit of the data was carried out using quantum Monte-Carlo \cite{Trotzky}. In Ref.~\cite{Derrico} and in Sec.~\ref{sec:phenomenological}, we pursue a  phenomenological approach to capture finite-temperature effects. 
Since temperature is expected to induce an exponential decay of the  correlations $g_{ij}$ at long distances $|i-j|$, the idea is to first do DMRG calculations at $T=0$, which are computationally cheap, and  to then multiply  the obtained correlators $g_{ij}(T=0)$  by $e^{-|i-j|/\xi_T}$.
The parameter $\xi_T$, in the following called \emph{effective thermal correlation length},  is left as the only free  parameter to fit the finite-$T$ experimental data.
Specifically, we introduce the modified correlations
\begin{equation}\label{eq:phenomAnsatz}
	\tilde{g}_{ij}(T) = C e^{-|i-j|/\xi_T}g_{ij}(T=0).
\end{equation}
The normalization factor $C$ is chosen such that the corresponding  momentum distribution $P(k)$ obeys $P(k=0)=1$. 
In the superfluid regime, this approach is motivated by Luttinger liquid theory~\cite{Giamarchibook}. 
In this theory the correlation function behaves as
\begin{equation}
g_{ij}(T) \propto \exp\left\{-\frac{1}{2K}\ln\left(\frac{\sinh(2K|i-j|/\tilde{\xi}_T)}{2Kd/n\tilde{\xi}_T}\right)\right\}
\label{eq:corr-finiteT}
\end{equation}
which interpolates between a power-law behavior when $|i-j| \ll\tilde{\xi}_T$ and an exponential behavior when $|i-j| \gtrsim \tilde{\xi}_T$.
Here $K$ is the dimensionless Luttinger parameter, which  is of order one in our case.
This formula is expected to be valid in the low-temperature regime with a thermal correlation length behaving as $\tilde{\xi}_T^{-1} = \frac{\pi}{2K}\frac{k_BT}{\hbar u}$, where $u$ is the sound velocity.
In the Luttinger liquid result \eqref{eq:corr-finiteT}, the exponential tail at finite $T$ is expected to depend on the particle density $n/d$. Hence, for inhomogeneous systems, one should rather have a site-dependent $\xi_T$, also varying from tube to tube. However, for the sake of simplicity, for each point in the diagram, we use a single $\xi_T$ for all tubes and all sites.

Of course, this approach is not exact and its validity depends on the temperature regime and the considered phase. It can be tested quickly on small homogeneous systems using exact diagonalization. Such a comparison shows that the phenomenological ansatz provides a sensible fit of the exact finite-$T$ data for the range of temperatures relevant for the experiment, i.e., $T \simeq J/k_B$.  The validity of the approach for the trapped system is discussed further in Sec.~\ref{sec:thermometry}. 

\subsection{Exact diagonalization for homogeneous systems}
\label{sec:methodED}
For small \emph{homogeneous} systems ($\alpha=0$), we use full diagonalization of the Hamiltonian \eqref{eq1} to obtain real-space correlations $g_{ij}$ at finite temperatures. Such correlation functions typically show an exponential decay that we fit using points with relative distance $\Delta z \leq 4d,5d$ to obtain the total correlation length $\xi(T)$. 
We use systems with various densities and sizes. Depending on the density, the system size $L$ ranges from $8d$ to $13d$.
Because of finite-size effects, the results are  useful  as long as $\xi(T)$ is sufficiently below the system size.

\subsection{Quasi-exact finite-\texorpdfstring{$T$}{T} DMRG computations}
\label{sec:methodDMRG-finiteT}
Zero-temperature DMRG computations \cite{White1992,White1993,Schollwock},
as employed in the approach described above, variationally optimize a certain ansatz for the many-body quantum state so-called matrix product states. While this only covers pure states, it can be extended to directly describe thermal states \cite{Verstraete,Feiguin,Barthel,Binder}. To this purpose, one computes a so-called purification of the thermal density matrix $\rho_{\beta}=e^{-\beta (H-\mu N)}$, where $\beta=1/k_B T$. Specifically, if the system is described by a Hilbert space $\mathcal{H}$, a purification $|\rho_{\beta}\rangle$ of the density matrix is a pure state from an enlarged Hilbert space $\mathcal{H}\otimes \mathcal{H}_\text{aux}$ such that $\rho_{\beta}=\operatorname{Tr}_\text{aux}|\rho_{\beta}\rangle\langle\rho_{\beta}|$, i.e., such that the density matrix is obtained by tracing out the auxiliary Hilbert space $\mathcal{H}_\text{aux}$  from the projector $|\rho_{\beta}\rangle\langle\rho_{\beta}|$. As the purification $|\rho_{\beta}\rangle$ is a pure many-body state, we can make a matrix product ansatz for it and deal with it in the framework of DMRG. Noting that it is simple to write down a purification for the infinite-temperature state $\rho_0=\mathbbm{1}$, one can start the computation at infinite temperature and use imaginary-time evolution to obtain finite-$T$ purifications $|\rho_\beta\rangle=e^{-\beta(H-\mu N)/2}\otimes\mathbbm{1}_\text{aux}|\rho_0\rangle$.
Based on this, finite-$T$ expectation values of any observable $A$ can be evaluated in the form $\langle A\rangle_\beta=\langle\rho_\beta|A\otimes\mathbbm{1}_\text{aux} |\rho_\beta\rangle/\langle\rho_\beta|\rho_\beta\rangle$.

\section{Phenomenological analysis of the \texorpdfstring{$U$-$\Delta$}{U-Delta} coherence diagram}
\label{sec:phenomenological}
\begin{figure}[t]
\centering
\includegraphics[width=\columnwidth] {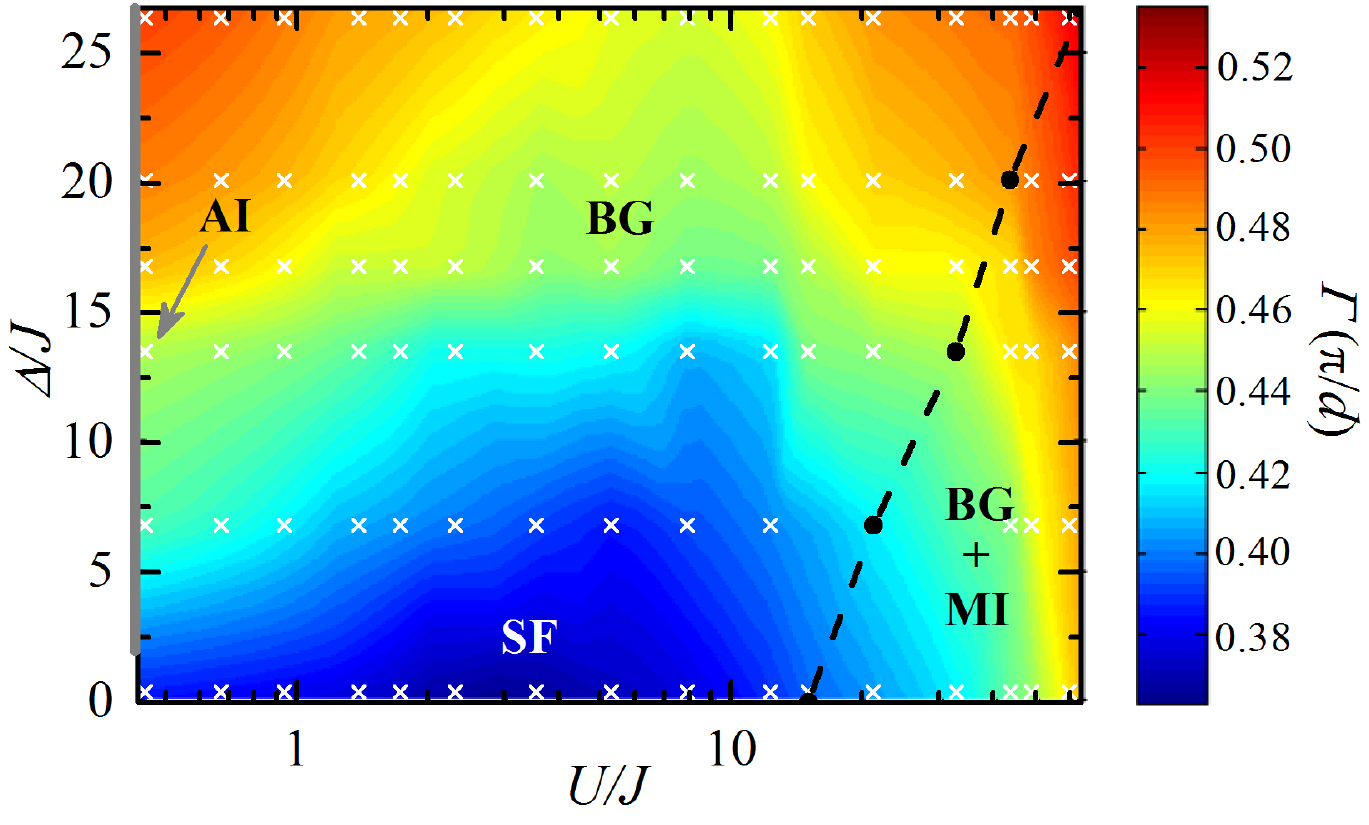}
\caption{Measured rms width $\Gamma$ of the momentum distribution $P(k)$ in the $U$-$\Delta$ diagram.  Without interaction ($U=0$), increasing $\Delta$ induces the transition from the superfluid (SF) to the Anderson insulator (AI). In the absence of disorder ($\Delta=0$), increasing $U$ leads to the superfluid-Mott insulator (MI) transition. For increasing $\Delta$ at large interaction, according to $T=0$ DMRG calculations, MI domains exist only at the right of the dashed line (i.e., $U>2\Delta$ for large $U$), where they coexist with SF or Bose glass (BG) domains, respectively below and above $\Delta=2J$. The diagram has been generated on the basis of 94 data points (crosses). Standard deviations of $\Gamma$ are between 2\% and 5\%. Data taken from Ref.~\cite{Derrico}.}
\label{fig:expdiagram}
\end{figure}
An overview of the insulating properties of our system is provided by measurements of the momentum distribution $P(k)$ \cite{Derrico}.  Obtained by interpolating 94 sets of measurements, Fig.~\ref{fig:expdiagram} shows the rms width $\Gamma$ of $P(k)$ as a function of the interaction strength $U$ and the disorder strength $\Delta$. The plot is representative of the phase changes occurring in the system. At small disorder and interaction values where the system is superfluid, $P(k)$ is narrow (blue zone). At larger disorder and interaction values, $P(k)$ progressively broadens (green, yellow, and red zones) meaning that the system is becoming more and more insulating.  
In particular, along the $\Delta=0$ line, the diagram is consistent with the progressive formation of a Mott insulator, which, in our inhomogeneous system, coexists with a superfluid fraction. For increasing $\Delta$ along the $U=0$ line, an Anderson insulator forms above the critical value $\Delta=2J$ predicted by the Aubry-Andr\'e model \cite{Roati08,Aubry}. For finite $U$ and $\Delta$, we observe a reentrant insulating regime extending from small $U$ and $\Delta>2J$ to large $U$, which surrounds a superfluid regime at moderate disorder and interaction. This shape is similar to that of the Bose glass phase found in theoretical studies of the $U$-$\Delta$ diagram for homogeneous systems at $T=0$ \cite{Giamarchi88, Fisher, Roux}.
\begin{figure}[t]
\centering
\includegraphics[width=0.9\columnwidth] {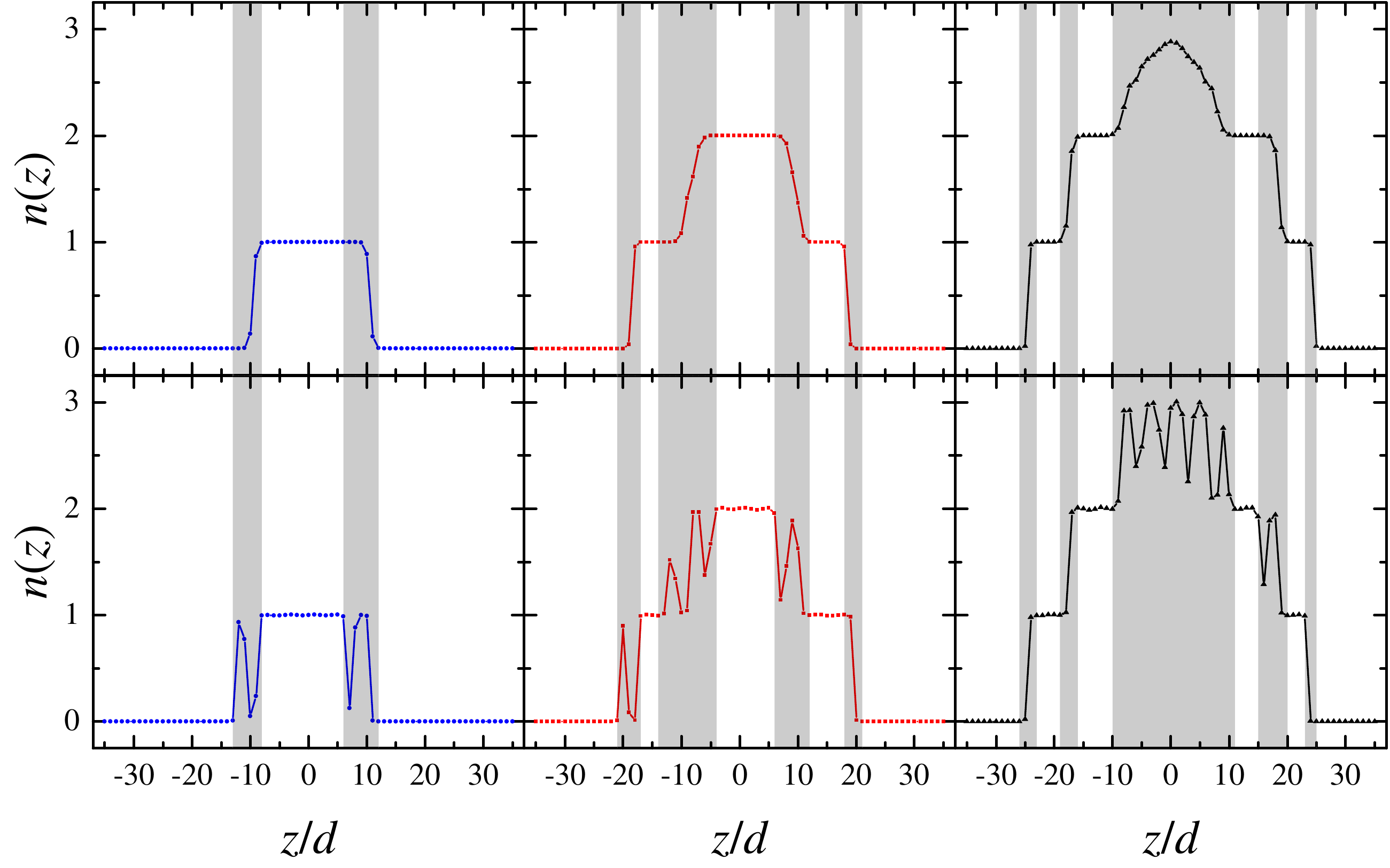}
\caption{Density profiles obtained from $T=0$ DMRG calculations for $U=26J$ and $\Delta=0$ (top) or $\Delta=6.5 J$ (bottom). Blue, red, and black curves refer to tubes with $N=20,55,96$ atoms, respectively. The shaded areas represent the regions with non-integer filling where the superfluid (top) becomes Bose glass (bottom). Data taken from supplemental material of Ref.~\cite{Derrico}.}
\label{fig:T0densityprofiles}
\end{figure}
\begin{figure*}[t]
\centering
\includegraphics[width=0.9\textwidth] {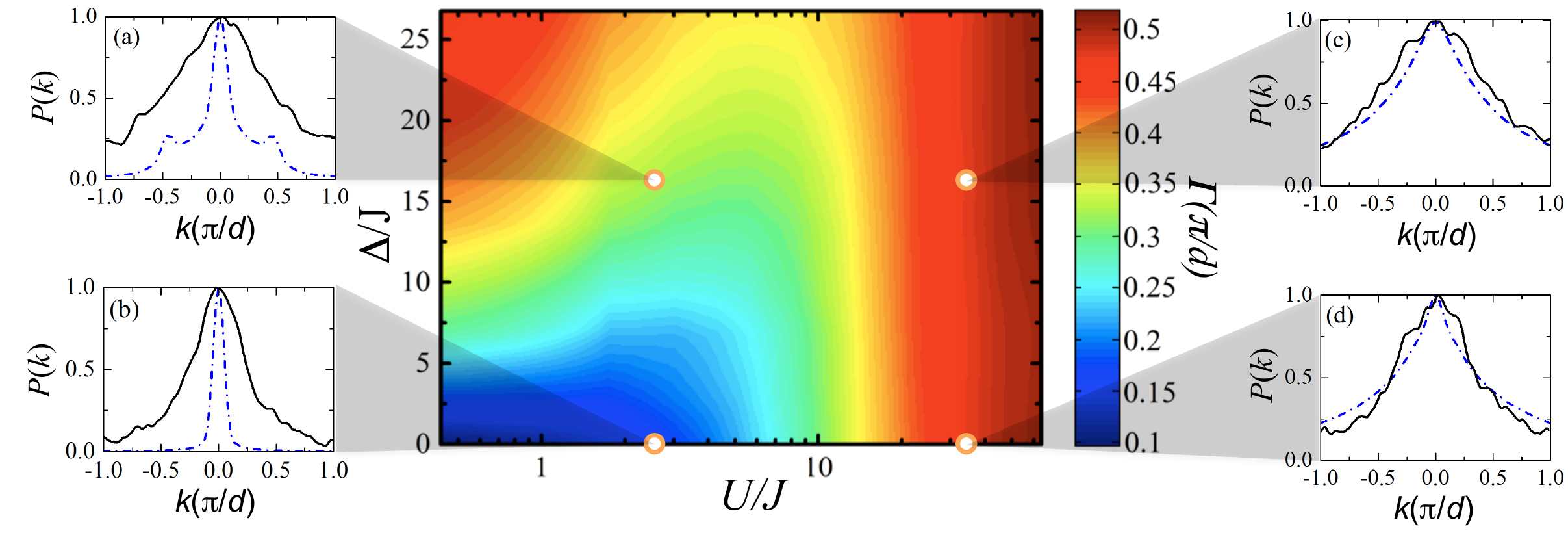}
\caption{Theoretical rms width $\Gamma$ of the  momentum distribution $P(k)$ at $T=0$, averaged over all tubes. The diagram is built from 94 data points as in the experimental diagram in Fig.~\ref{fig:expdiagram}. For few representative points, $P(k)$ is also shown at the side of the diagram: the theoretical result for $T=0$ (blue, dot-dashed) is compared to the experimental finite-$T$   data (black, solid). Data taken from Ref.~\cite{Derrico}.} 
\label{fig:T0PD}
\end{figure*}

The coexistence of different phases due to the trapping potential can be observed clearly in density profiles, which can be computed numerically by DMRG. For example, Fig.~\ref{fig:T0densityprofiles} gives the calculated density profiles for $T=0$ in tubes with $N=20, 55, 96$ atoms in the strong-interaction regime.
For these strong interactions and in the absence of disorder (top), the profiles show the typical wedding cake structure, where the commensurate Mott domains (integer $n$) are separated by incommensurate superfluid regions (non-integer $n$). Adding disorder (bottom), the Mott regions progressively shrink and the smooth density profiles of the incommensurate regions become strongly irregular, as expected in the case of a Bose glass.
Note that the dashed line in Fig.~\ref{fig:expdiagram} delimits the region of the diagram where Mott-insulating domains appear at zero-temperature. These domains are quantitatively defined by the condition that, in the $T=0$ DMRG density profiles for the three representative tubes with $N=20, 55, 96$ atoms, there are at least three consecutive sites with integer filling. 

The challenge of the investigation of the experimental diagram and of its comparison with the ideal theoretical case lies in the inhomogeneity and in the finite temperature, especially, as the temperature is not directly accessible in the experiment. 
In the following, we first compare the experimental finite-$T$ diagram with DMRG calculations reproducing our inhomogeneous system at $T=0$. Subsequently, a phenomenological extension of the $T=0$ results to finite temperatures provides a more quantitative understanding of the temperature-induced coherence loss.

\subsection{Zero-temperature \texorpdfstring{$U$-$\Delta$}{U-Delta} diagram}

Let us theoretically study the behavior of the momentum distribution $P(k)$ [Eq.~\eqref{eq:MomentumDistribution}] of the model \eqref{eq1}.
Fig.~\ref{fig:T0PD} shows the full $U$-$\Delta$ coherence diagram at $T=0$  in terms of the rms width $\Gamma$ of $P(k)$, together with a few distributions $P(k)$ at representative points. The data are based on  the TF hypothesis for the distribution of particles among tubes. Indeed, using the GC hypothesis would  require to compute all $N(\mu)$ curves across the diagram which is rather expensive numerically. In contrast to the typical phase diagrams for homogeneous systems \cite{Roux}, here, only crossovers between regimes occur, as different phases can coexist due to the inhomogeneity of the system. Still, Fig.~\ref{fig:T0PD} shows the same three main regions occurring in the experimental diagram; in particular, the strongly-correlated regime for large interaction strengths with a reentrance of the localization. 
However, the different ranges of the color scales reveal the quantitative difference between the theoretical $T=0$ results and the experimental finite-$T$ results in Fig.~\ref{fig:expdiagram}. 
In particular, for small $U$ (left panels in Fig.~\ref{fig:T0PD}), the numerical $T=0$ momentum distributions (blue, dot-dashed curves) are considerably narrower than the experimental finite-$T$ ones (black, solid curves). Conversely, for large $U$ (right panels), the thermal broadening is much less relevant. In the following, we try to better understand and quantify this aspect using first the phenomenological approach.

\subsection{Phenomenological approach and elementary interpretation of the coherence diagram}
\label{sec:phenomenologicalapproach}
\begin{figure*}[t]
\centering
\includegraphics[width=0.9\textwidth] {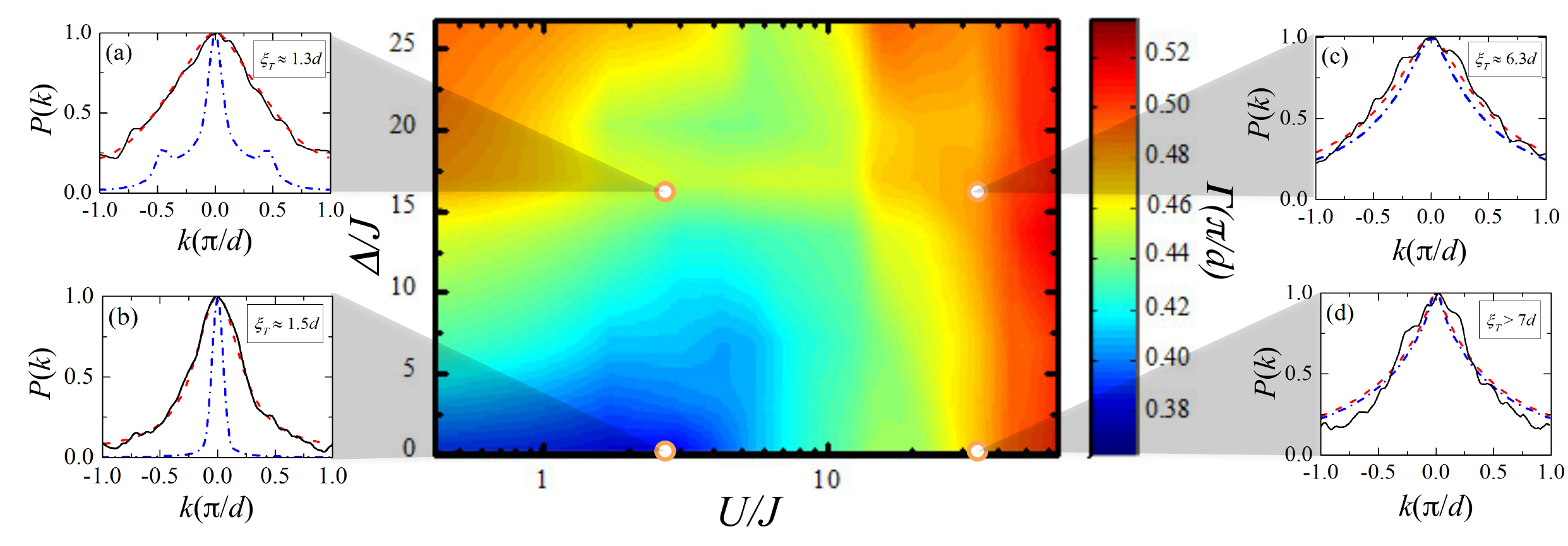}
\caption{$U$-$\Delta$  diagram for the rms width $\Gamma$ of the phenomenological $P(k)$ (red, dashed), obtained as the convolved momentum distribution (see text) that fits the experimental $P(k)$ (black, solid). 
The thermal correlation length $\xi_T$  is the fitting parameter that phenomenologically accounts for thermal effects according to the ansatz \eqref{eq:phenomAnsatz}. The full diagram is generated by interpolation from the same $U$-$\Delta$ points as in Fig.~\ref{fig:expdiagram}. Data taken from  Ref.~\cite{Derrico}.} 
\label{fig:phenomenologicalPD}
\end{figure*}
\begin{figure}[b]
\centering
\includegraphics[width=\columnwidth] {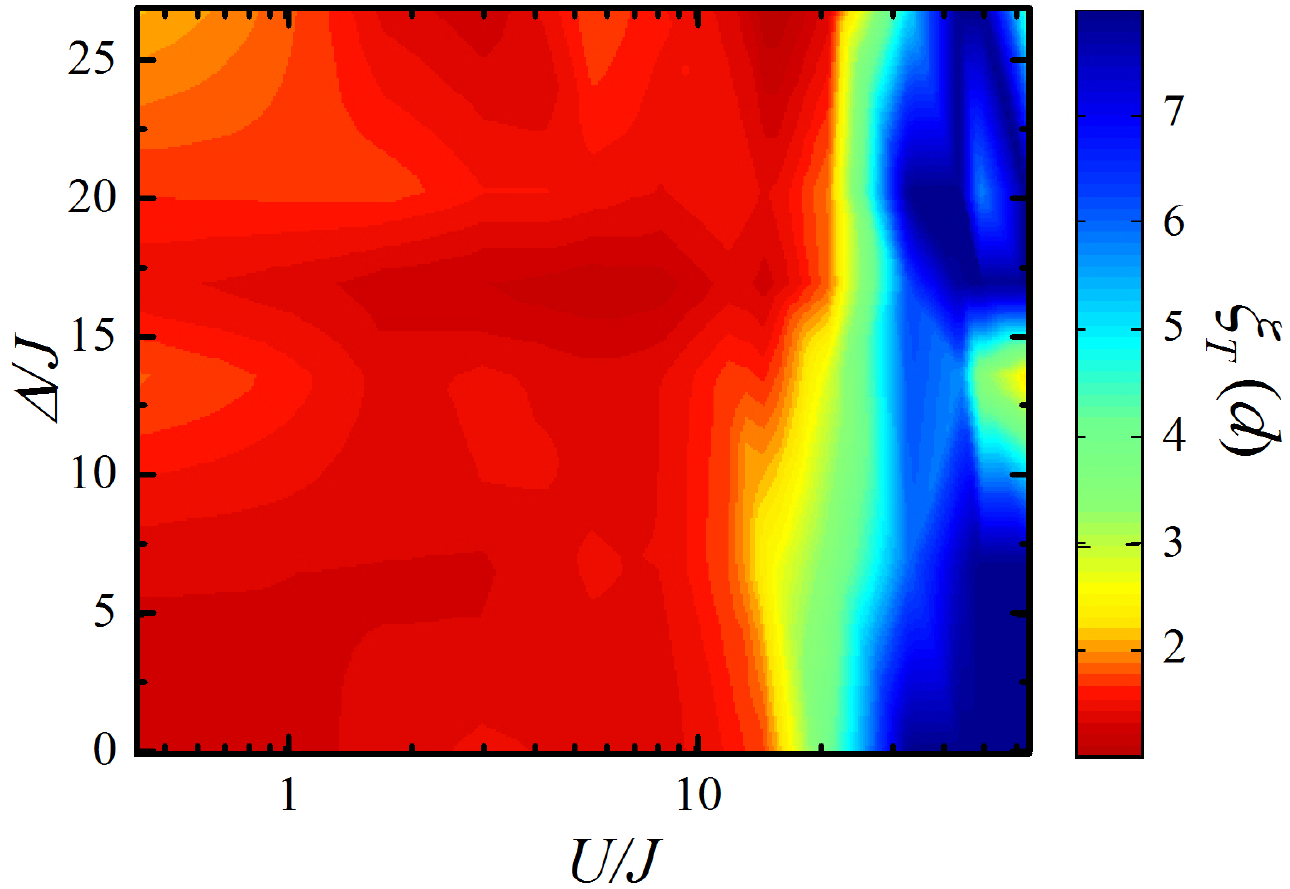}
\caption{$U$-$\Delta$ diagram of the thermal correlation length $\xi_T$ resulting from the  phenomenological ansatz (\ref{eq:phenomAnsatz}) by fitting it to the experimental momentum distribution $P(k)$.  Thermal effects are significantly more relevant for small $U$.  Data taken from supplemental material of Ref.~\cite{Derrico}.}
\label{fig:phenomenologyXiT}
\end{figure}
A natural source of broadening of the momentum distribution $P(k)$ is the temperature, and we address its effect for the whole $U$-$\Delta$ diagram based on the phenomenological approach explained above in Sec.~\ref{sec:phenomAnsatz}. The phenomenological approach has the advantages of simplicity and of a direct connection to the described $T=0$ results with TF distribution of atoms among tubes, yielding a first elementary interpretation for temperature effects.

For each point in the diagram, we systematically fitted the experimental distribution $P(k)$ with the phenomenological ansatz resulting from Eq.~\eqref{eq:phenomAnsatz}, leaving the effective thermal correlation length $\xi_T$ as a single fit parameter. Some typical fits (red, dashed curves) are shown in side panels of Fig.~\ref{fig:phenomenologicalPD}. The main part of the figure shows the rms width $\Gamma$ of the phenomenological momentum distribution across the whole $U$-$\Delta$ diagram. This should be compared to the experimental diagram in Fig.~\ref{fig:expdiagram}, employing the same color scale.
  The obtained $\Gamma$ values are similar across the whole diagram, except for the large-$U$ and small-$\Delta$ region, where the fits are not good.
As  explained in the next section, this discrepancy is due to the completely different thermal response of the coexisting superfluid and Mott-insulating components. 

A rough interpretation of the diagram is that the inverse total correlation length, denoted by $\xi(T)$, is approximately given by the sum of the inverses of an intrinsic ($T=0$) correlation length, denoted by $\xi_0$, and the thermal correlation length $\xi_T$.  This is summarized by the formula
\begin{equation} 
\label{eq:totxi}
	\frac{1}{\xi(T)} = \frac{1}{\xi_0} + \frac{1}{\xi_T}.
\end{equation}
The zero-temperature correlation length $\xi_0$ is finite in the localized Mott-insulating and Bose glass regimes. In homogeneous systems $\xi_0$ diverges in the superfluid regime and $\xi(T)$ would then be identical to the effective thermal correlation length $\xi_T$. For our inhomogeneous systems, $\xi_0$ becomes large in the superfluid regime, but remains finite.

We can interpret $\xi_T$ as a quantification of the thermal broadening which is obtained, according to Eq.~\eqref{eq:phenomAnsatz}, by convolving the theoretical zero-temperature momentum distribution $P(k)$ of width  $1/\xi_0$ with a Lorentzian distribution of width $1/\xi_T$. Depending on the point in the diagram, one may then separate the intrinsic zero-temperature and the thermal contributions to the observed broadening. Remember that both $\xi_0$ and $\xi_T$ are \emph{effective} correlation lengths appearing after averaging over many inhomogeneous tubes and are in principle not directly related to the correlation lengths for a homogeneous system, although they are expected to follow the same trends with interaction and disorder. 

The behavior of $\xi_T$ as extracted from the fits is shown in Fig.~\ref{fig:phenomenologyXiT} for the whole $U$-$\Delta$ diagram. For $U<10J$, $\xi_T$ is rather short, $d\lesssim \xi_T\lesssim 2d$, showing that thermal broadening is important for the superfluid and weakly interacting Bose glass regimes. Moreover, $\xi_T$ does not strongly vary as a function of $\Delta$. This shows that the overall increase of $\Gamma$  with increasing $\Delta$ in Fig.~\ref{fig:phenomenologicalPD}  is essentially due to a decrease of the intrinsic correlation length $\xi_0$. In this context it is important to note that, when increasing $\Delta$, the localization length in the considered quasi-periodic model \eqref{eq1} can reach  values much smaller than the lattice spacing $d$ more rapidly than in the case of  true random potentials~\cite{Roux}. This is favorable for the experiment, which then probes the strongly localized Bose glass regime. In the superfluid region, the thermal contribution to the broadening is clearly dominating and the observed small values of $\xi_T$ correspond to a gas with short-range quantum coherence.

Let us now discuss the large-$U$ regime. There, the obtained $\xi_T$ are significantly larger, suggesting that the strongly correlated phases are only weakly affected by finite-temperature effects. For large $U$ in the Mott phase,  $\xi_0$ can get much smaller than $d$. Here, the rms width is dominated by the intrinsic $T=0$ width, as confirmed directly by the fits in the side panels of Fig~\ref{fig:phenomenologicalPD}. Importantly, this shows that the observed reentrance of the localization in the experimental diagram is driven by interactions and disorder, and not by thermal effects. 
\begin{figure}[t]
\centering
\includegraphics[width=0.95\columnwidth] {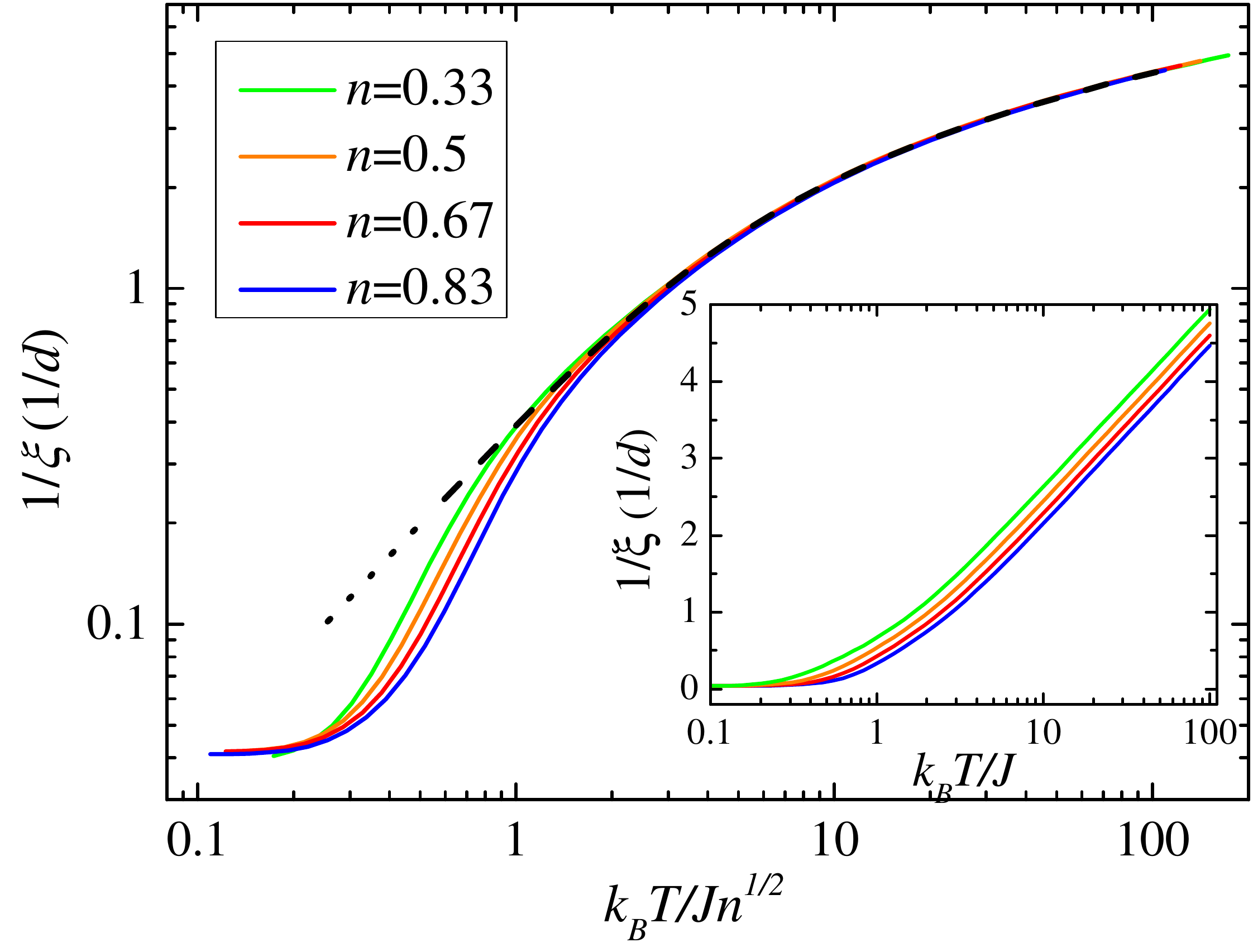}
\caption{Temperature dependence of the inverse correlation length $1/\xi(T)$, calculated by exact diagonalization for a small system ($L=12d$) in the superfluid regime ($\Delta=0$, $U=2J$), for various site occupancies $n \leq 1$. The dashed line is a fit of the numerical high-$T$ data with Eq.~\eqref{xi}. Inset: density dependence of $1/\xi(T)$. As shown in the main panel, the change of density $n$ can be taken into account  by a scaling factor such that, when $\xi(T)$ is plotted versus $k_BT/Jn^{1/2}$, all curves overlap for $k_BT\gtrsim 2J$. Data taken from supplemental material of Ref.~\cite{Derrico}.}
\label{fig:EDSFregime}
\end{figure}

\section{Effect of temperature on the correlation length from exact diagonalization}
\label{sec:ED}
As the effective thermal correlation lengths $\xi_T$  in the phenomenological approach are found to be relatively short with respect to the system size, one can gain a first understanding of the temperature dependence of the correlation length $\xi(T)$  from exact diagonalization calculations for small-sized systems, as described in Sec.~\ref{sec:methodED}. Let us stress that the validity of this analysis is limited to the regions of the phase diagram, where the correlation length $\xi(T)$ is sufficiently shorter than the considered system sizes $L\in[8d,12d]$.

\subsection{Thermal broadening for weak interactions}
\label{sec:EDSF}
Fig.~\ref{fig:EDSFregime} shows the temperature dependence of the inverse correlation length $\xi^{-1}(T)$ at $U=2J$ (superfluid regime) for several densities below $n=1$. The data show a crossover from a low-temperature regime $k_B T \ll J$ to a high-temperature regime $k_B T \gg J$. When $U$ is not too large, a natural energy scale is set by the bandwidth $4J$ that controls this crossover. With exact diagonalization, we cannot investigate the low-temperature regime due to finite-size effects, but let us recall that, according to the Luttinger liquid field theory~\cite{Giamarchibook}, a linear behavior $\xi^{-1} \sim k_BT/Jd$ is expected, with a prefactor that depends on density and interactions. In the opposite regime of high $T$, we are able to determine the correct scaling of the correlation length from exact diagonalization. In the range $2J\lesssim k_BT\lesssim 100J$, which is also the range of experimental interest, the numerical results are very well fitted by the relation
\begin{equation} \label{xi}
	\xi^{-1}(T) \simeq d^{-1}{\rm arcsinh}\left(\frac{k_B T}{c J\sqrt{n}}\right)
\end{equation}
with $c=2.50(5)$ being a fit parameter, valid for the relevant range of densities and interaction $U=2J$. This formula is inspired by the one given in Ref.~\cite{Sirker} for free fermions, $\xi(T)\simeq d/{\rm arcsinh}(k_BT/J)$. For high temperatures, $\xi^{-1}(T)$ is thus logarithmic in $T$, corresponding to a ``classical'' limit of the lattice model and is attributed to the finite bandwidth.
We do not have a theoretical argument for the observed $\sqrt{n}$ scaling; so it should be taken as an ansatz that describes the data for the given values of $U$ and $n$, but not as a general formula.
\begin{figure}[t]
\centering
\includegraphics[width=0.9\columnwidth] {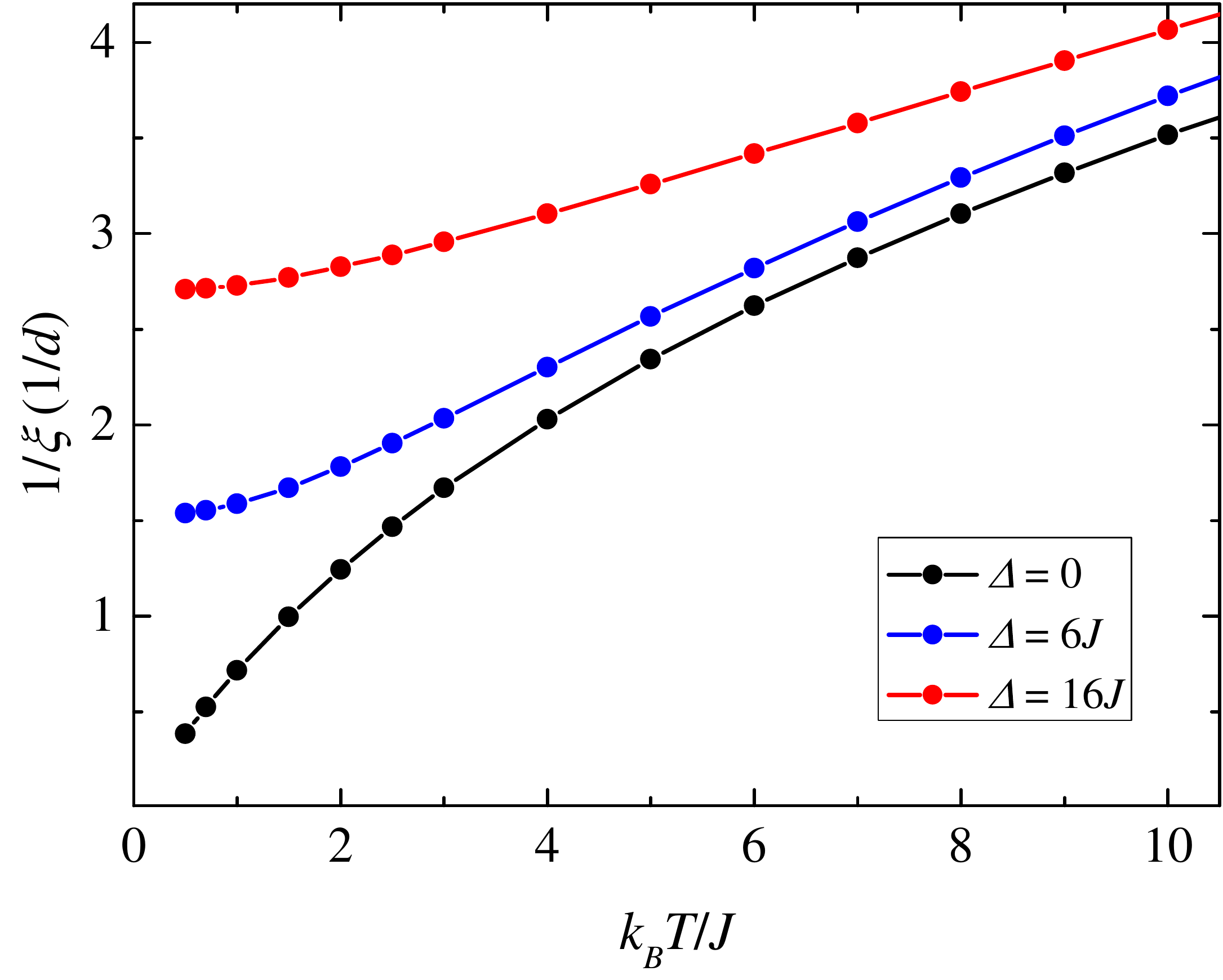}
\caption{Temperature dependence of the inverse correlation length $1/\xi(T)$, calculated by exact diagonalization and derived from a Lorentzian fit of $P(k)$ for a weakly interacting system ($U=2.3J$) with $L=13d$ and $n=0.46$, for three values of the disorder strength: $\Delta=0, 6J, 16J$.}
\label{figS6}
\end{figure}

Additional computations performed in the presence of disorder (see Fig.~\ref{figS6}) confirm the previous results of strong thermal effects for small $U$. For disordered systems, fluctuations of  the local density and hence of the correlation functions  are much more relevant. Thus, in small sized-systems, trying to fit the  exponential decay of the  real-space correlations proves to be difficult.  We hence determine  the inverse correlation length $\xi^{-1}(T)$ from a Lorentzian fit of the momentum distribution $P(k)$.
$\xi^{-1}(T)$ starts to increase at rather small $T$, showing that there is a non-negligible impact of thermal fluctuations already at low temperatures. This explains the  short $\xi_T$ observed in the analysis of the experimental data for weak interactions.
It is however interesting to point out that, according to recent studies on transport properties of the same system, the broadening of $P(k)$ with $T$ is not accompanied by a change of the system mobility \cite{Derrico}. Further investigations of this persisting insulating behavior at finite $T$ might establish a link with the many-body localization problem \cite{Aleiner,Michal}.

\subsection{Quantum-normal crossover temperature for strong interaction}
\label{crossoverT}
\begin{figure}[t]
\centering
\includegraphics[width=0.9\columnwidth] {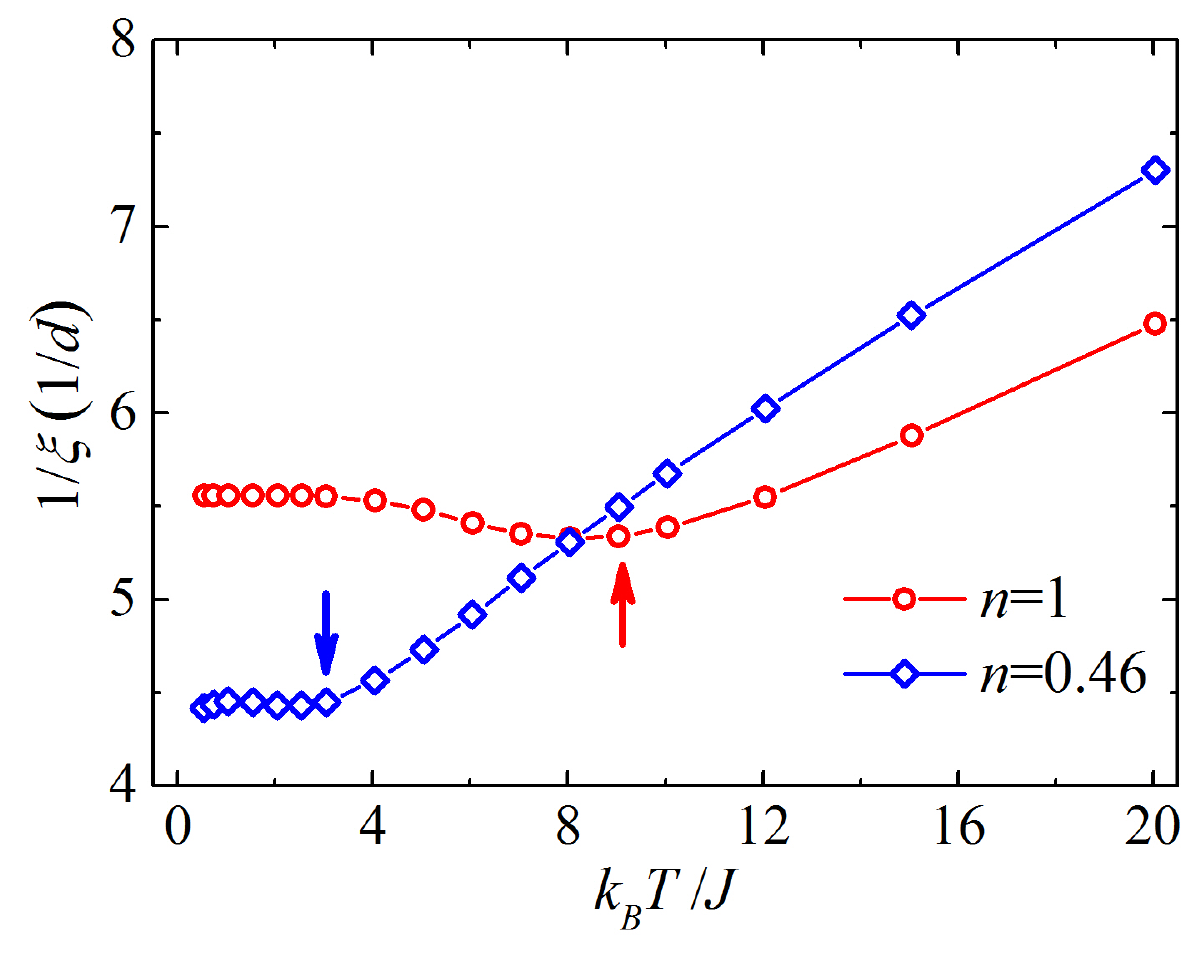}
\caption{Temperature dependence of the inverse correlation length, calculated by exact diagonalization for a strongly interacting system with $U=44J$, disorder strength $\Delta=10J$ and for both the commensurate case of a Mott insulator ($n$=1) and the incommensurate case of a Bose glass ($n$=0.46). The system lengths are $L=9d$ and $L=13d$, respectively. Arrows indicate the crossover temperatures $T_0$ below which $\xi^{-1}(T)$ is rather constant before starting increasing.}
 \label{fig8_}
\end{figure}

Let us now discuss the temperature dependence of the correlation length for strong interactions ($U>10J$). As shown in Fig.~\ref{fig8_}, $\xi(T)$ is only weakly dependent on $T$ at low temperatures while a relevant broadening sets in above a crossover temperature $T_0$. This effect can be seen clearly not only for the Mott phase, for which it occurs when the thermal energy becomes comparable with the energy gap $U$ \cite{Gerbier}, but also for the gapless Bose glass. 
$T_0$ is here determined as the position of the maximum of the derivative of $1/\xi(T)$.

\begin{figure}[t]
\centering
\includegraphics[width=0.9\columnwidth] {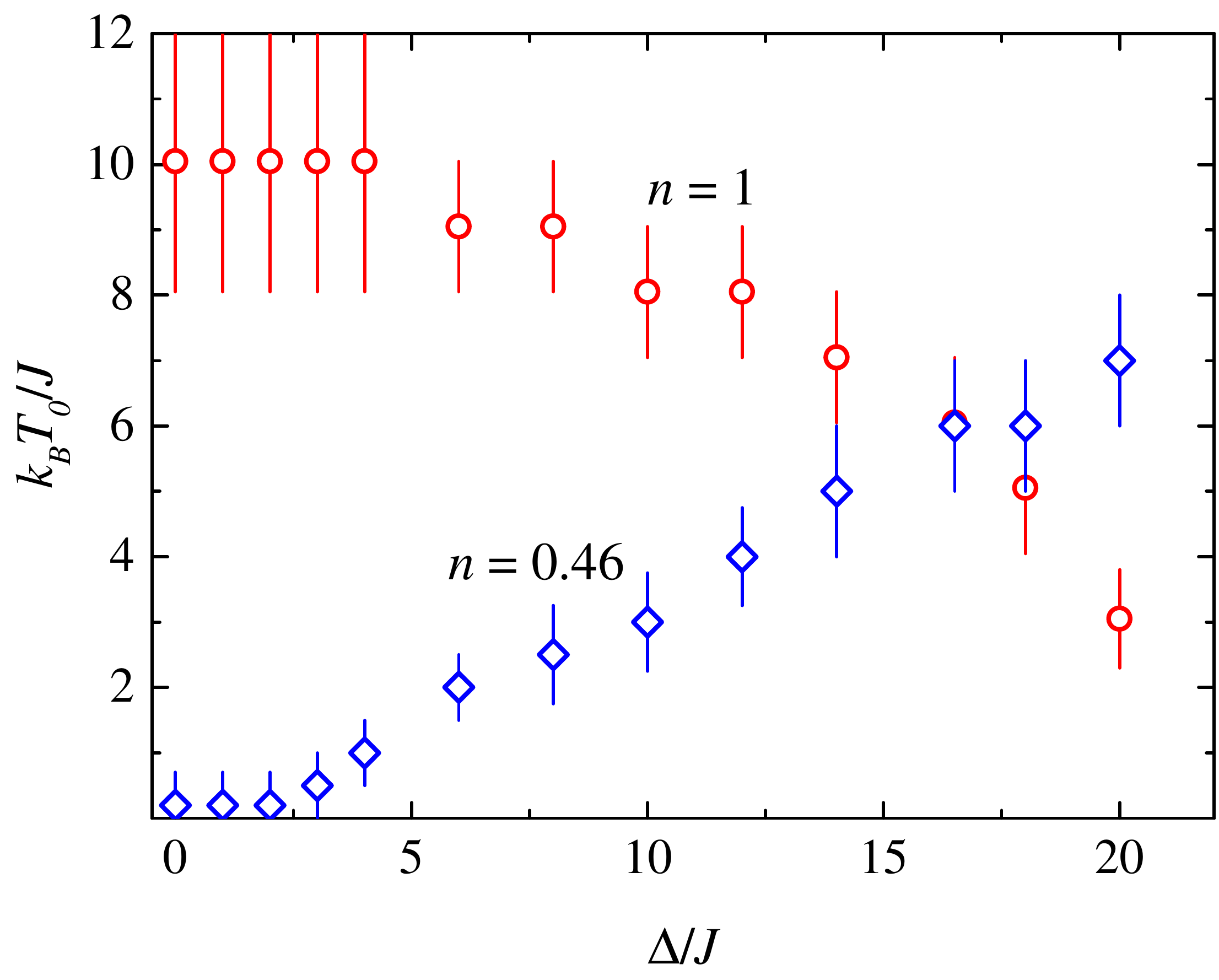}
\caption{Crossover temperature $T_0$ as a function of the disorder strength $\Delta$, calculated by exact diagonalization for a strongly interacting system with $U=44J$, for both the commensurate case of a Mott insulator ($n$=1) and the incommensurate case of a Bose glass ($n$=0.46). The system lengths are $L=9d$ and $L=13d$, respectively.}
\label{figS7}
\end{figure}

Fig.~\ref{figS7} shows the computed crossover temperature $T_0$ as a function of the disorder strength $\Delta$ for a representative interaction strength and for both a commensurate  and an incommensurate density.
For the commensurate density and $\Delta=0$, we obtain $k_BT_0=0.23(6)U$, which is comparable to the Mott insulator ``melting'' temperature $k_B T\simeq 0.2U$, predicted for higher-dimensional systems \cite{Gerbier}. As $\Delta$ increases, $T_0$ decreases, which is consistent with a reduction of the gap due to the disorder.
For the Bose glass (incommensurate density), the crossover temperature shows instead a linear increase with $\Delta$, i.e., $k_BT_0\propto\Delta$. This result, already observed in numerical simulations at small disorder strengths \cite{Nessi}, can be intuitively justified with the following reasoning. The energies of the lowest levels that the fermionized bosons can occupy increase with the disorder strength. So the larger  $\Delta$, the higher the effective Fermi energy that sets the temperature scale for the existence of the quantum phase (the Bose glass).

The exact diagonalization results confirm those obtained in Sec.~\ref{sec:phenomenological} with the phenomenological approach: we showed in Fig.~\ref{figS7} that for sufficiently large $\Delta$, $\xi(T)$ is not significantly affected by the finite temperature. This is in agreement with the large $\xi_T$ obtained phenomenologically (see Fig.~\ref{fig:phenomenologyXiT} at large $U$).

Finally, the fact that the crossover temperatures in the incommensurate and commensurate cases are different for small disorder and strong interaction, clarifies why in the phenomenological approach the fit of the momentum distribution with a single $\xi_T$ is not working properly in this regime, as previously mentioned. In particular, while the superfluid component  broadens in the same way as it does for small $U$, the weakly-disordered Mott-insulating component for $T<T_0$ does not. As a consequence, considering a single thermal broadening leads to an overall overestimation of the derived $\Gamma$.

\subsection{Experimental momentum width versus entropy}
\label{sec:GammaVsS}

Since a procedure to determine the experimental temperature in a disordered system is not available, a direct comparison of theory and experiment is not possible. Nevertheless we can give a first experimental indication of the consistency of the previous results by investigating the correlation length  as a function of  entropy, which we can measure as described below. 
\begin{figure}[ht]
\centering
\includegraphics[width=0.85\columnwidth] {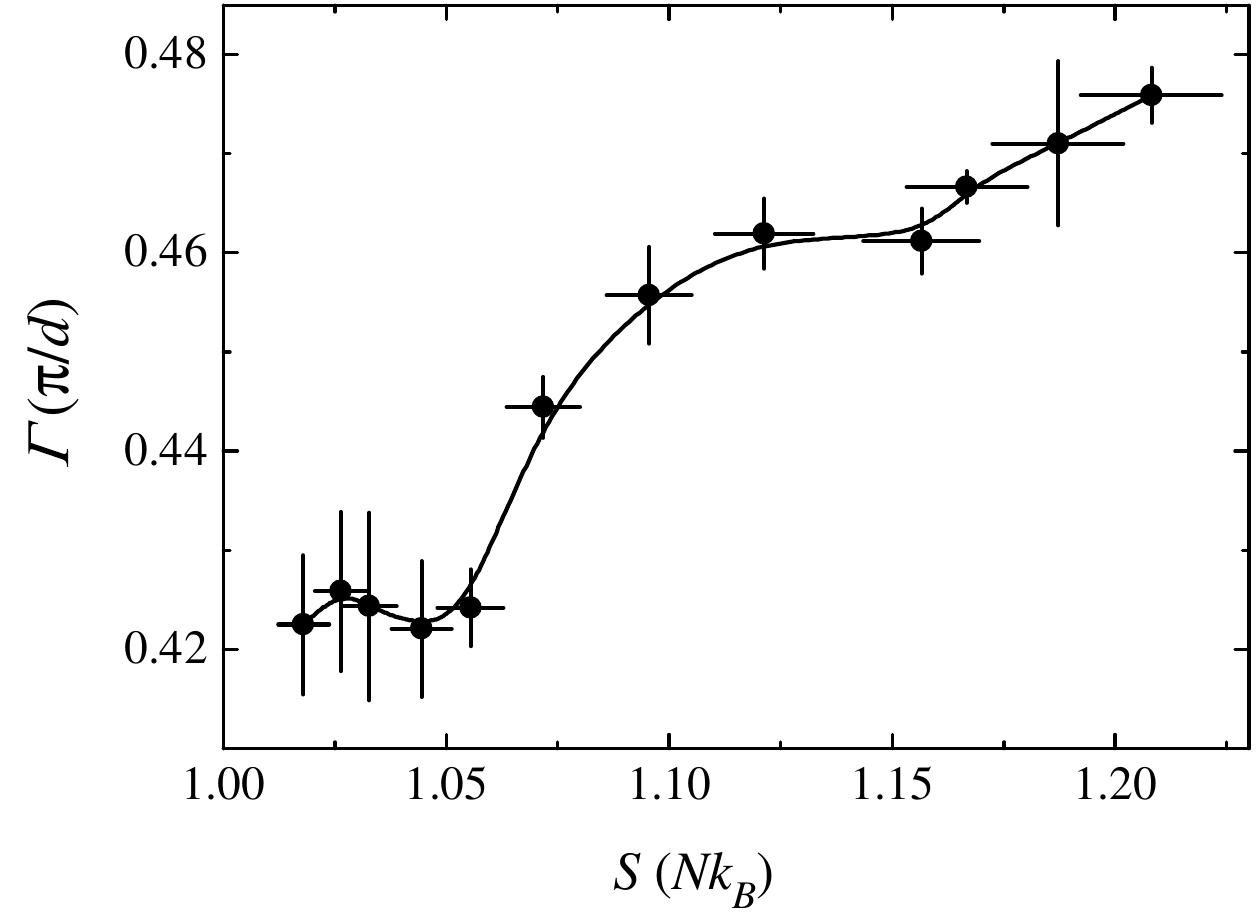}
\caption{Measured rms width $\Gamma$ of the momentum distribution $P(k)$ for $U=23.4J$ and $\Delta=6.6J$ as a function of the estimated entropy per particle. $\Gamma$ starts increasing only above a certain entropy value. The uncertainties are the standard deviation of typically five measurements. The line is a guide to the eye.}
\label{fig11}
\end{figure}

In Fig.~\ref{fig11}, we report the measured rms width $\Gamma$ of the momentum distribution $P(k)$ as a function of the entropy  in the regime of strong interaction and finite disorder, where the Bose glass and the disordered Mott-insulating phases coexist. 
The measurement clearly shows the existence of a plateau at low entropy, before a broadening sets in, which nicely recalls  the theoretical behavior of the inverse $\xi(T)$ in Fig.~\ref{fig8_}. Assuming a monotonic increase of temperature with entropy, this experimental result supports the theoretical prediction that, for sufficiently large $U$ and $\Delta$, the $T=0$ quantum phases can persist in the finite-$T$ experiment.

The entropy in the 1D tubes is estimated as follows. We first measure the initial entropy of the system in the 3D trap: in the BEC regime with $T/T_c <1$, where $T_c$ is the critical temperature for condensation in 3D, we use the relation $S=4N_T k_B \zeta(4)/\zeta(3)(T/T_c)^3$, where $\zeta$ is the Riemann Zeta function \cite{Catani}. The reduced temperature $T/T_c$ is estimated from the measured condensed fraction by taking into account the finite interaction energy. 
 After slowly ramping the lattices up and setting the desired values of $U$ and $\Delta$, we again slowly ramp the lattices down,  such that only the 3D trapping potential remains, and we again measure the entropy as just described. As an estimation for the entropy in the 1D tubes we use the mean value of these initial and final entropies. Through variation of the waiting time, the amount of heating can be changed and we can hence obtain the rms width $\Gamma$ for different entropies.

The data in the experimental coherence diagram, Fig.~\ref{fig:expdiagram}, and the lowest-entropy point of Fig.~\ref{fig11} correspond to the shortest used waiting times.
For example, the lowest-entropy point in Fig.~\ref{fig11} has the same rms width ($\Gamma \simeq 0.42 \pi/d$) as the one obtained for the coherence diagram at $U=23.4J$ and $\Delta=6.6J$.

\section{Thermometry with finite-\texorpdfstring{$T$}{T} DMRG}
\label{sec:thermometry}

A standard procedure for obtaining the temperature of a quasi-condensate in a harmonic trap is to use the linear relation  $T= \hbar^2 n\, \delta p/0.64k_B m d$  between the temperature $T$ and the half width at half maximum $\delta p$ of the Lorentzian function that fits the experimental momentum distribution \cite{Gerbier03,  Gerbier04}. 
However, so far, there exists no formula for the temperature in the interacting disordered or clean lattice systems. Here, we perform \emph{ab initio} finite-$T$ DMRG computations of $P(k)$ to estimate $T$, both in the superfluid and Mott-insulating regimes. 

We note that in Ref.~\cite{Derrico} we actually provided a rough estimate of the superfluid temperature, $T\simeq3J/k_B$. The value was obtained by inverting Eq.~\eqref{xi} for the approximate temperature dependence of the correlation length $\xi(T)$ and replacing $\xi(T)$ by  the effective thermal correlation length  $\xi_T$  obtained with the phenomenological approach.
(According to Eq.~\eqref{eq:totxi}, in the superfluid regime $\xi(T)\approx\xi_T$  as $\xi_0$ is considerably larger than $\xi_T$).
In this simplified approach, the inhomogeneity of the system  was taken into account by performing a local density approximation (LDA). The more precise finite-$T$ DMRG analysis, described in the following, yields a superfluid temperature that is twice as large as the old estimate.

As described in Sec.~\ref{sec:methodDMRG-finiteT}, using finite-$T$ DMRG, we can perform  \emph{ab initio} calculations to obtain $P(k)$.  This allows for a proper thermometry of the system and also for testing the validity of the phenomenological approach. After quasi-exact simulation of the system for different temperatures, the resulting momentum distributions $P(k)$ are compared with the experimental data to estimate the experimental temperature.
We restrict the analysis to two points on the $\Delta=0$ axis of the diagram: one for $U= 3.5J$, corresponding to the superfluid regime, and another one for $U= 21J$, which is deeply in the strong-interaction regime. Let us recall the general trends for the rms width $\Gamma$ of $P(k)$: $\Gamma$ typically increases with the interaction strength $U$ and the temperature $T$, and also when the number of particles $N_\nu$ in a tube is decreased. As the momentum distribution is normalized to $P(k=0)=1$, low-filled tubes display flat tails while highly filled ones yield a more peaked momentum distribution. Lastly, one should keep in mind  that the exact distribution of atoms among the tubes in the experiment is not known. We therefore study both the Thomas-Fermi (TF) and grand-canonical (GC) hypotheses as described in Sec.~\ref{sec:distrib}.
\begin{figure}[t]
\centering
\includegraphics[width=0.9\columnwidth] {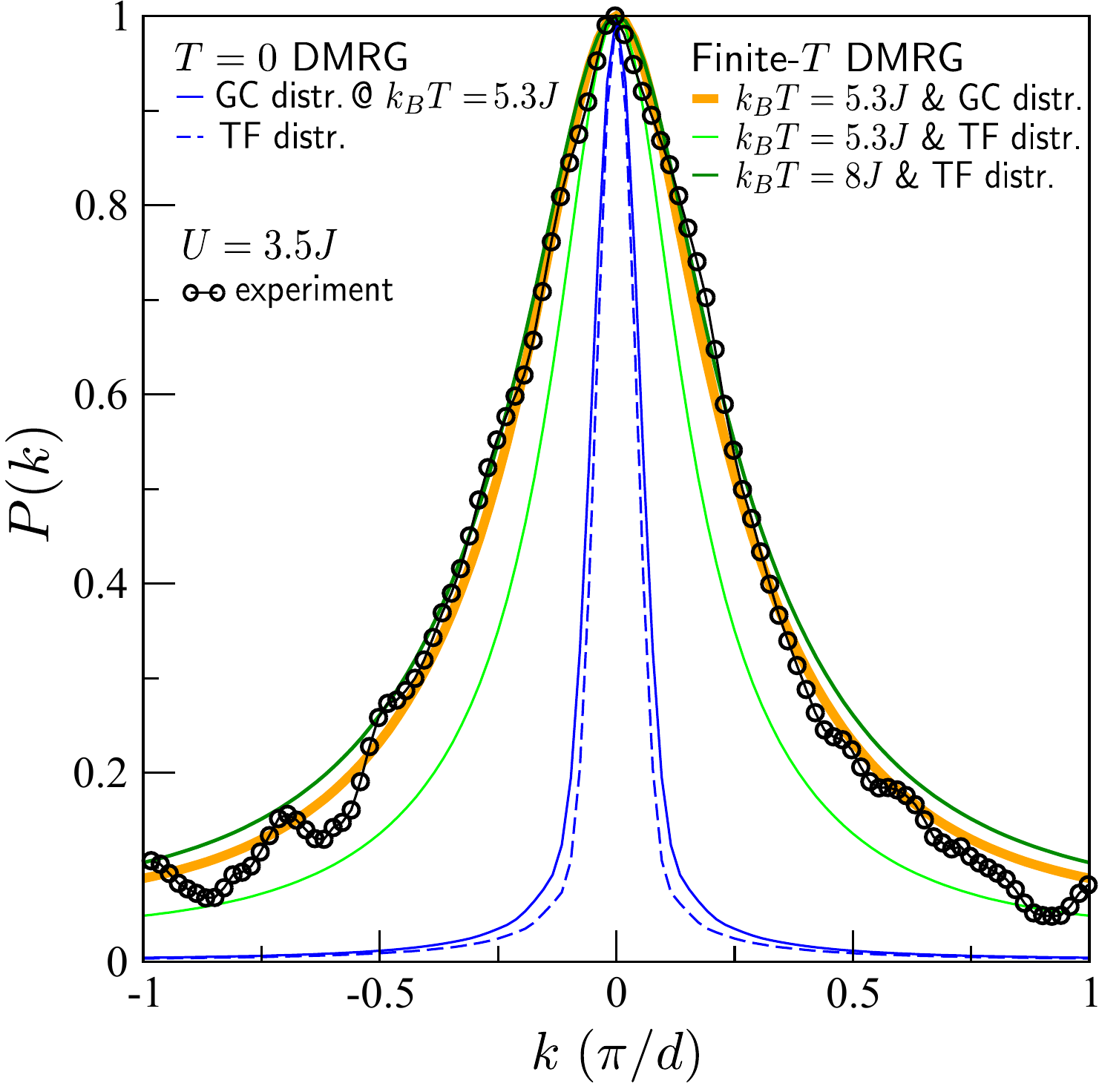}
\caption{Thermometry in the superfluid regime ($U=3.5J$) from finite-$T$ DMRG calculations for both the hypotheses of a grand-canonical (GC) or Thomas-Fermi (TF) distribution of the particles among the tubes.}
\label{fig:TDMRGsuperfluid}
\end{figure}

\subsection{In the superfluid regime}
\begin{figure}[b]
\centering
\includegraphics[width=0.55\columnwidth] {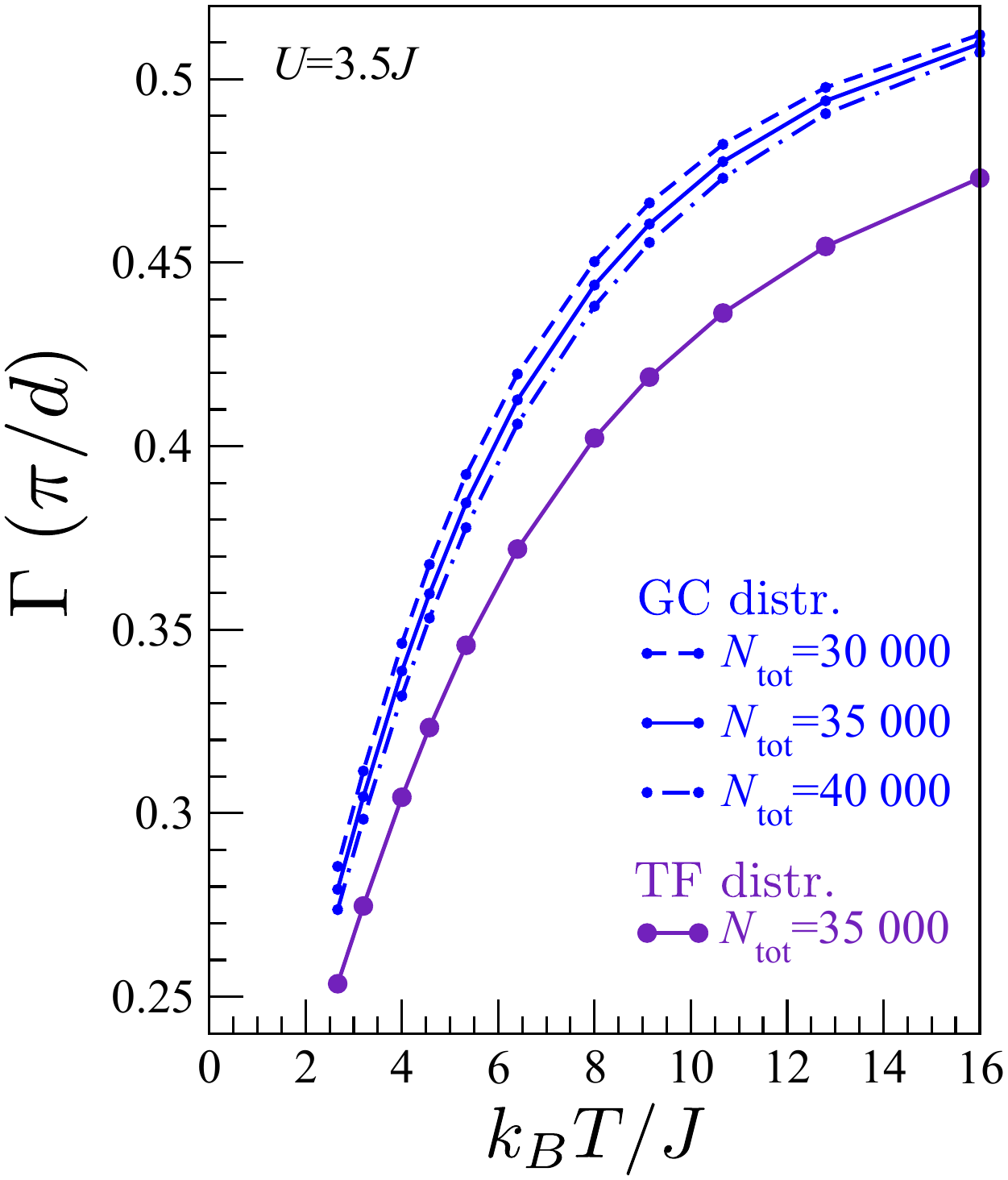}
\caption{Temperature dependence of the rms width $\Gamma$ of the momentum distribution $P(k)$ in the superfluid regime from finite-$T$ DMRG calculations under both the GC and TF assumptions. In the former case, the effect of varying the total number of particles $N_{\text{tot}}$ is also shown.}
\label{fig:TDMRGGammaPlot}
\end{figure}
\begin{figure}[t]
\centering
\includegraphics[width=0.9\columnwidth] {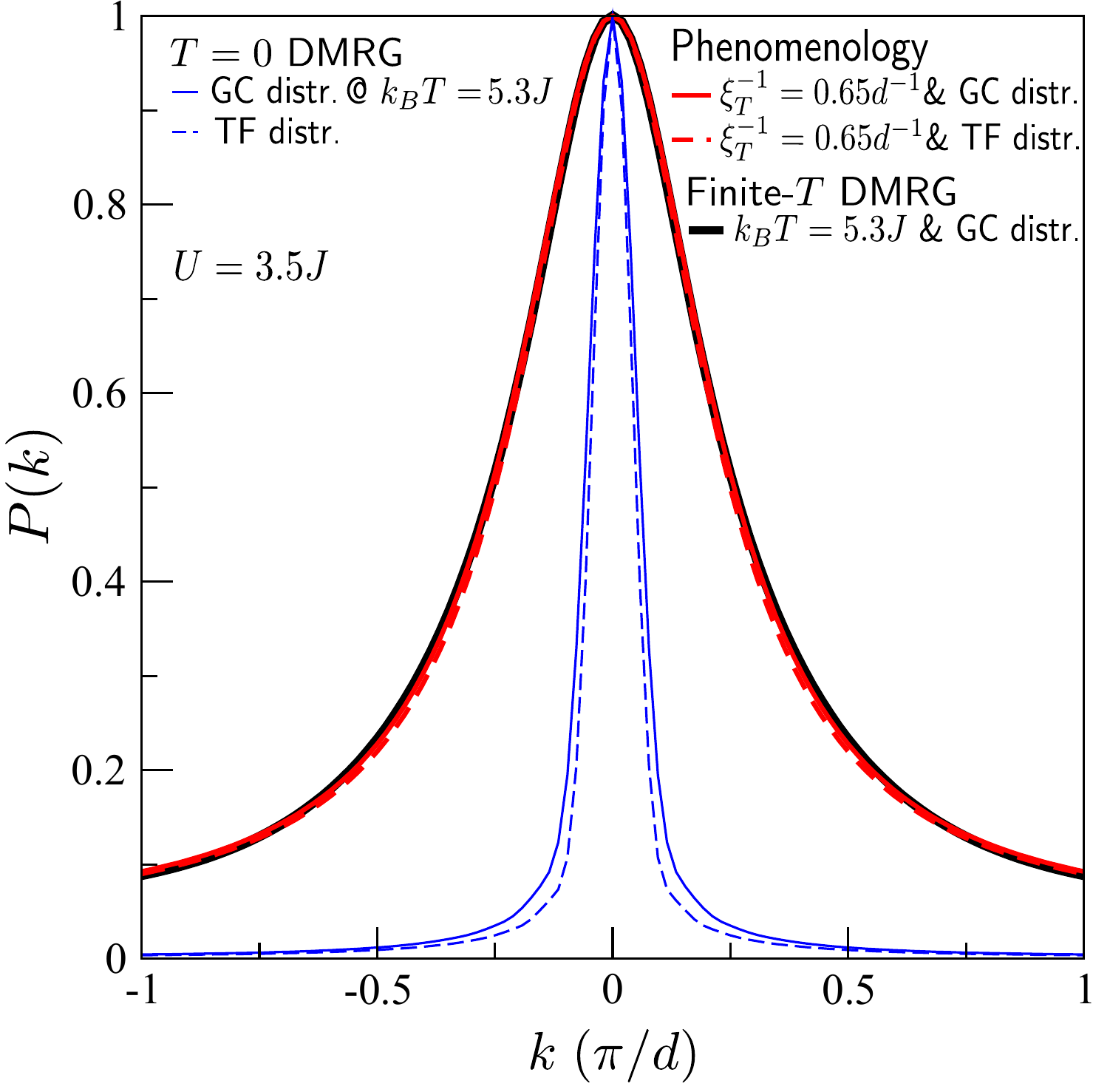}
\caption{Testing the phenomenological approach \eqref{eq:phenomAnsatz} for $U=3.5J$ and $T=5.3J/k_B$. To this purpose, $T=0$ DMRG data for the momentum distribution are folded with a Lorentzian that corresponds to the effective thermal correlation length $\xi_T=d/0.65$. The results are compared with the quasi-exact finite-$T$ DMRG data.}
\label{fig:TDMRGphenomenology}
\end{figure}
In Fig~\ref{fig:TDMRGsuperfluid} we show the results for the superfluid regime.
To estimate the temperature from experimental data we compute several theoretical $P(k)$ curves for different temperatures  and select the one that best matches the experimental $P(k)$. For the chosen interaction strength $U=3.5J$, a good estimate for the temperature is found to be $T=5.3J/k_B$ assuming the GC distribution for particles among tubes (bold orange curve in Fig.~\ref{fig:TDMRGsuperfluid}). The theoretical result matches the experimental data rather well, except for some oscillations in the tails that are due to correlated noise from the apparatus. Fig.~\ref{fig:TDMRGsuperfluid} also shows the theoretical $P(k)$ under the hypothesis of the TF distribution of the particles with temperatures $T=5.3J/k_B$ and $T=8J/k_B$. The former is more peaked and hence less wider than the GC curve for the same temperature. The latter is the best fit of the experimental data under the TF hypothesis. With the TF hypothesis we thus obtain larger temperature estimates than with the GC one. This is consistent with the general dependence of $P(k)$ on the particle number $N$ and the particle number distributions. As shown in Fig.~\ref{fig:NofMu},  the GC distribution has more particles in outer low-filled tubes and less in the higher-filled inner ones, when compared to the TF distribution.
To show that thermal broadening is certainly relevant in the considered parameter regime, Fig.~\ref{fig:TDMRGsuperfluid} also shows the narrow $P(k)$, obtained from $T=0$ DMRG data for both the TF distribution and the GC one for $T=5.3J/k_B$.

In Fig.~\ref{fig:TDMRGGammaPlot} we report  the  rms width $\Gamma$ of the momentum distribution $P(k)$ as a function of temperature, as obtained by finite-$T$ DMRG computations, for both the GC and TF distributions. It shows that, for temperature estimates, the knowledge about atom distribution is more important than the present 15\% fluctuations in the number of atoms $N_{\text{tot}}$.  
As the GC approach takes into account a possible redistribution of the atoms among tubes during the slow ramping of the lattice potentials, we consider it to be more realistic and reliable than the TF one, which, in a sense,  freezes the particle distribution to that in the initial 3D trap. 

As already mentioned, the temperatures obtained with $T$-DMRG ($T=5.3J/k_B$ with the GC approach and $T=8J/k_B$ with the TF one) are higher with respect to the rough estimate ($T\simeq3J/k_B$) presented in Ref.~\cite{Derrico}, where we performed exact diagonalization calculations with a LDA.
 Yet, the order of magnitude is the same. The finite-$T$ DMRG approach is in principle much more reliable as it is basically approximation-free and takes into account the actual system sizes and trapping potentials.  While exact diagonalization results, combined with LDA, do not take into account properly the system inhomogeneity, they can nevertheless easily provide the general trend of the correlation length with temperature.\\

With the exact finite-$T$ calculations, we can also test the phenomenological approach discussed for the full coherence diagram in Sec.~\ref{sec:phenomenological}. 
For both the TF and GC hypotheses,  in Fig.~\ref{fig:TDMRGphenomenology}, we show the data for $T=0$ DMRG (blue) and for the phenomenological approach with $\xi_T^{-1}=0.65d$ (red). The latter curves are compared to actual $T=5.3J/k_B$ finite-$T$ DMRG data under the GC hypothesis (black).
The agreement is rather good for both the TF and GC distributions since the corresponding $T=0$ curves for $P(k)$ are already similar. It is interesting to note that, despite the inhomogeneity of the system, assuming a single effective thermal correlation length $\xi_T$ in the phenomenological approach  [Eq.~\eqref{eq:phenomAnsatz}]  works nicely in the superfluid regime, where the rms width $\Gamma$ is dominated by thermal broadening.

While the phenomenological approach, based on $T=0$ DMRG data and on the effective thermal correlation length $\xi_T$, here yields the correct functional form for the thermal $P(k)$, it does not allow to determine the temperature precisely. The temperature dependence of $\xi_T$ can be obtained rather well from exact diagonalization for homogeneous systems as long as  $T$ is not too low (cf.\ Sec.~\ref{sec:EDSF}). However, its dependence on the atom distribution is not so easy to predict. So, for the phenomenological approach, the difficulty lies in the fact that very similar $P(k)$ can be obtained with the two considered particle distributions at quite different temperatures as documented by the exact results in Fig.~\ref{fig:TDMRGsuperfluid}.

\subsection{In the Mott-insulating regime}
\begin{figure}[t]
\centering
\includegraphics[width=0.9\columnwidth] {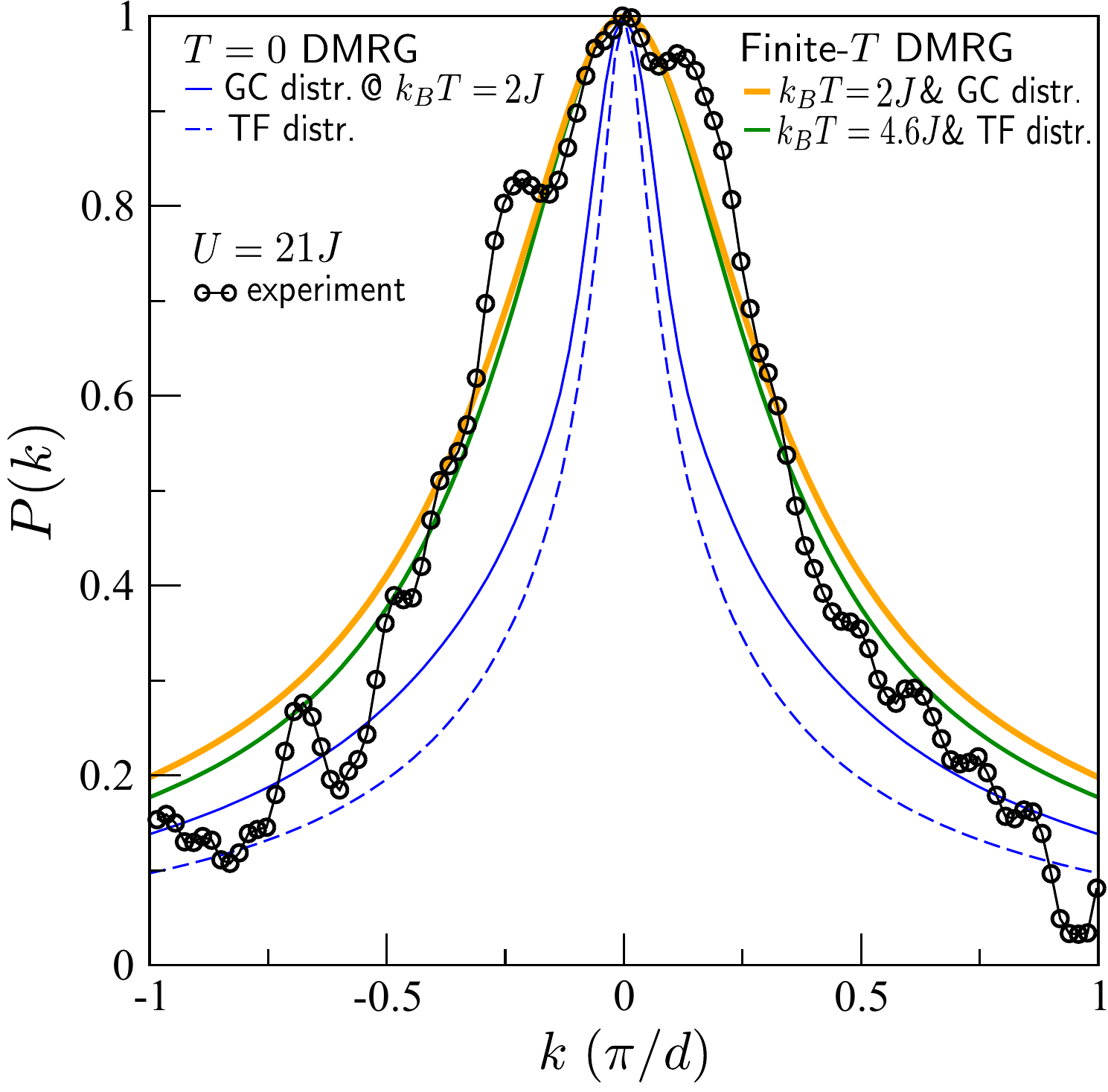}
\caption{Thermometry in the strong-interaction regime ($U=21J$) from finite-$T$ DMRG calculations, for both the hypotheses of a GC or TF distribution of the particles among the tubes.}
\label{fig:TDMRGmott}
\end{figure}
\begin{figure}[t]
\centering
\includegraphics[width=0.9\columnwidth] {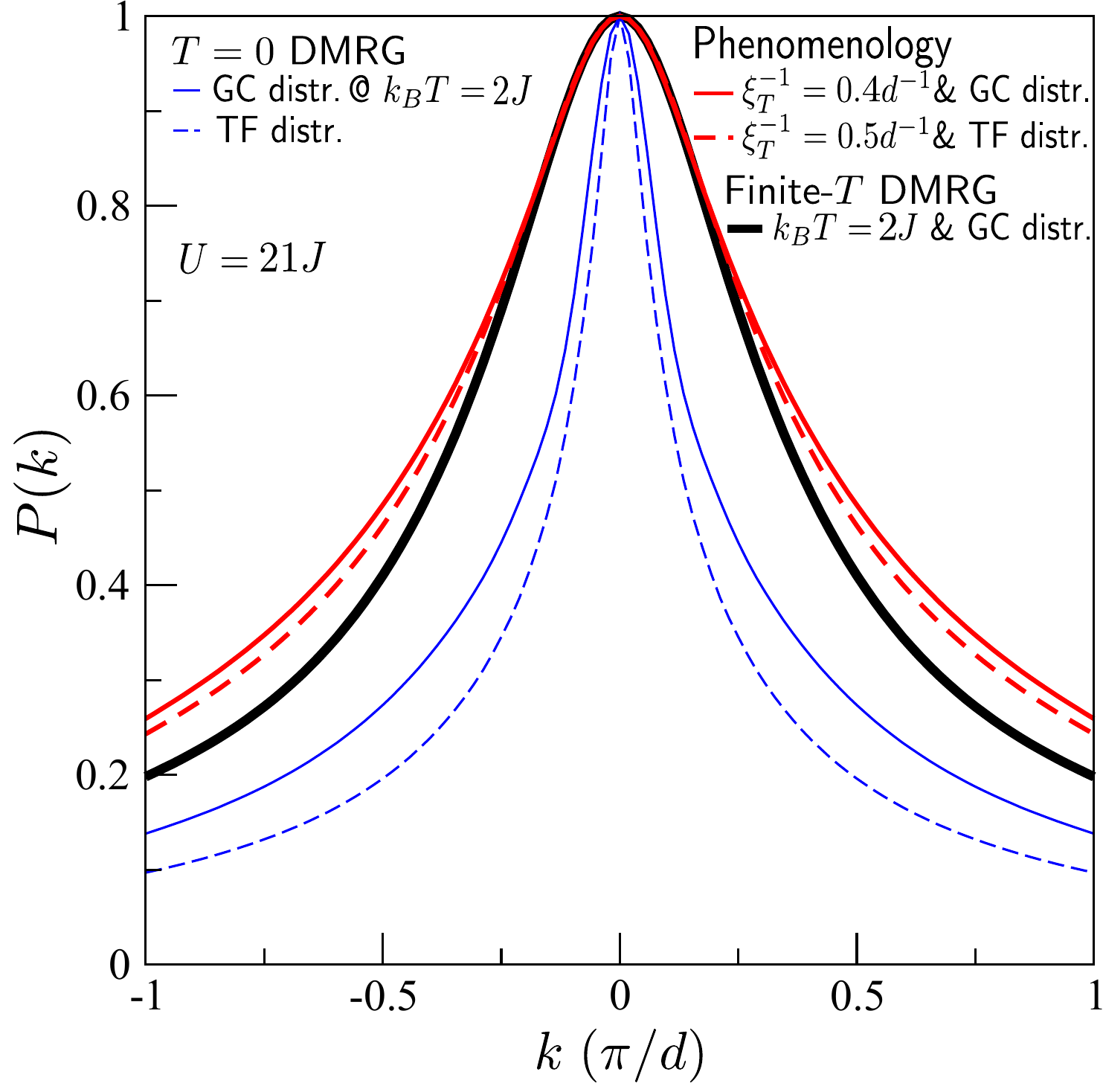}
\caption{Testing the phenomenological approach \eqref{eq:phenomAnsatz}  for $U=21J$ and  $T=2J/k_B$. To this purpose, $T=0$ DMRG data for the momentum distribution are folded with a Lorentzian that corresponds to thermal correlation lengths $\xi_T=d/0.4$ and $\xi_T=d/0.5$, respectively for the GC and TF cases. The results are compared with the quasi-exact finite-$T$ DMRG data.}
\label{fig:TDMRGphenomott}
\end{figure}

Similar comparisons are  carried out for the strong-interaction regime with $U= 21J$. The data are shown in Figs.~\ref{fig:TDMRGmott} and \ref{fig:TDMRGphenomott}.
For larger $U$ values the momentum distributions $P(k)$ for a single tube are typically wider. Yet, for such a tube with $T=0$, the rms width is not a monotonous function of the number of particles because of the wedding cake structure. For instance, particles added to a Mott plateau in the bulk will eventually form a superfluid dome that will contribute with a narrower signal to the $P(k)$ curve of the tube. Consequently, at low temperatures, this regime is  more sensitive to the particle distribution than the superfluid one. This is already visible in  the $T=0$ data for the TF and GC hypotheses. 
Contrary to the superfluid regime, the matching of  the theoretical curves (GC and TF)  with the experimental one is not very convincing since one cannot account equally well for  the central dome and the tails of the momentum distribution at the same time. As a rough estimate for the temperature we obtain $T\approx 2 J/k_B$ under the GC hypothesis. As in the superfluid case, this is smaller than the value ($4.6 J/k_B$) obtained under the TF hypothesis. In any case, experimental temperatures in the Mott regime are apparently lower than those in the superfluid regime.

The discrepancy between theory and experiment should be mainly due to thermalization issues.  In the inhomogeneous system experimental temperatures could vary spatially, since the insulating components, which are less susceptible to heating  because of the Mott gap, do not thermalize with the superfluid components~\cite{Delande, Bernier}.\\
 
As done in the superfluid case, we can again use finite-$T$ DMRG to test the phenomenological approach (cf.\ Fig.~\ref{fig:TDMRGphenomott}). 
The phenomenological ansatz for the momentum distribution, corresponding to Eq.~\eqref{eq:phenomAnsatz}, is fitted to exact DMRG data for $T=2J/k_B$.
The effective thermal correlation lengths $\xi_T$ are chosen to best fit the central dome of the exact curve, 
although this results in considerable deviations in the tails. Such deviations are however in agreement with the fact that, as already explained in Sec.~\ref{crossoverT}, the commensurate component of the Mott insulator thermally broadens less than the incommensurate superfluid one, leading to an overestimation in the phenomenological broadening of the tails.

\begin{figure}[t]
\centering
\includegraphics[width=0.55\columnwidth] {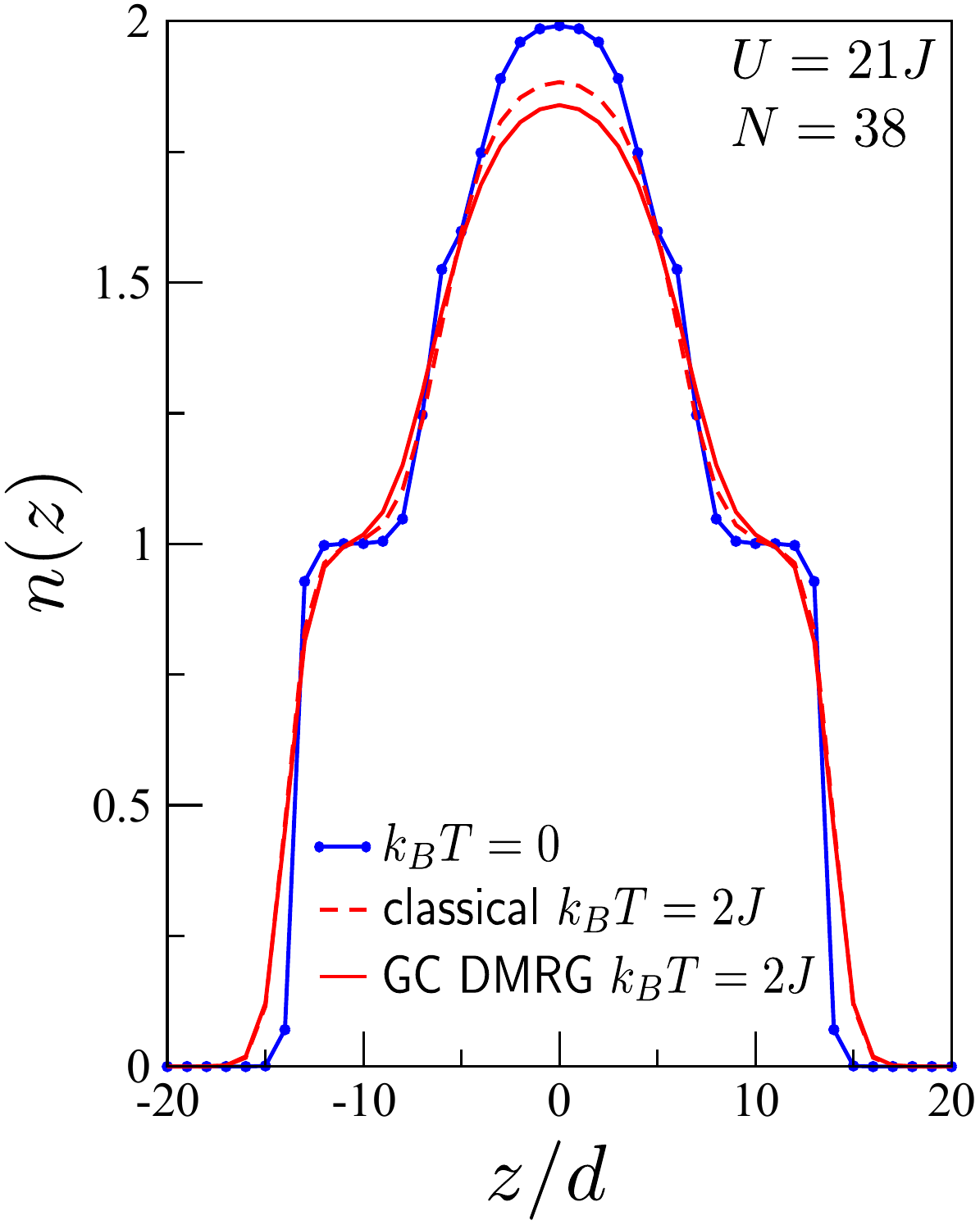}
\caption{ DMRG density profile in a tube at $T=0$ and at the estimated experimental temperature  for $U=21J$. Comparison with the classical model is also given.}
\label{fig:profilesU21}
\end{figure}

An additional complication originates from the fact that finite-size systems are more sensitive to temperature. At $T= 2J/k_B$, the shortest Mott plateaus, like for example those shown in Fig.~\ref{fig:profilesU21}, have almost completely melted despite the fact that the aforementioned estimate $T \simeq 0.2U/k_B$ for the melting temperature yields $4J/k_B$ at this interaction strength. 
 This means that the $T=0$ correlation functions, employed for the phenomenological approach, differ qualitatively from the actual finite-$T$ correlations.

\section{Experimental \texorpdfstring{$U$-$\Delta$}{U-Delta} entropy diagram}
\label{sec:experimentaltemperatures}

Thermometry on the basis of finite-$T$ DMRG in principle allows to also determine the system temperature  in the presence of disorder. 
However, to get  reliable temperature estimates one should ensure that the experimental system is in thermal equilibrium. 
As discussed previously, thermalization is hampered by localization in the Mott insulator and Bose glass phases.
\begin{figure}[t]
\centering
\includegraphics[width=\columnwidth] {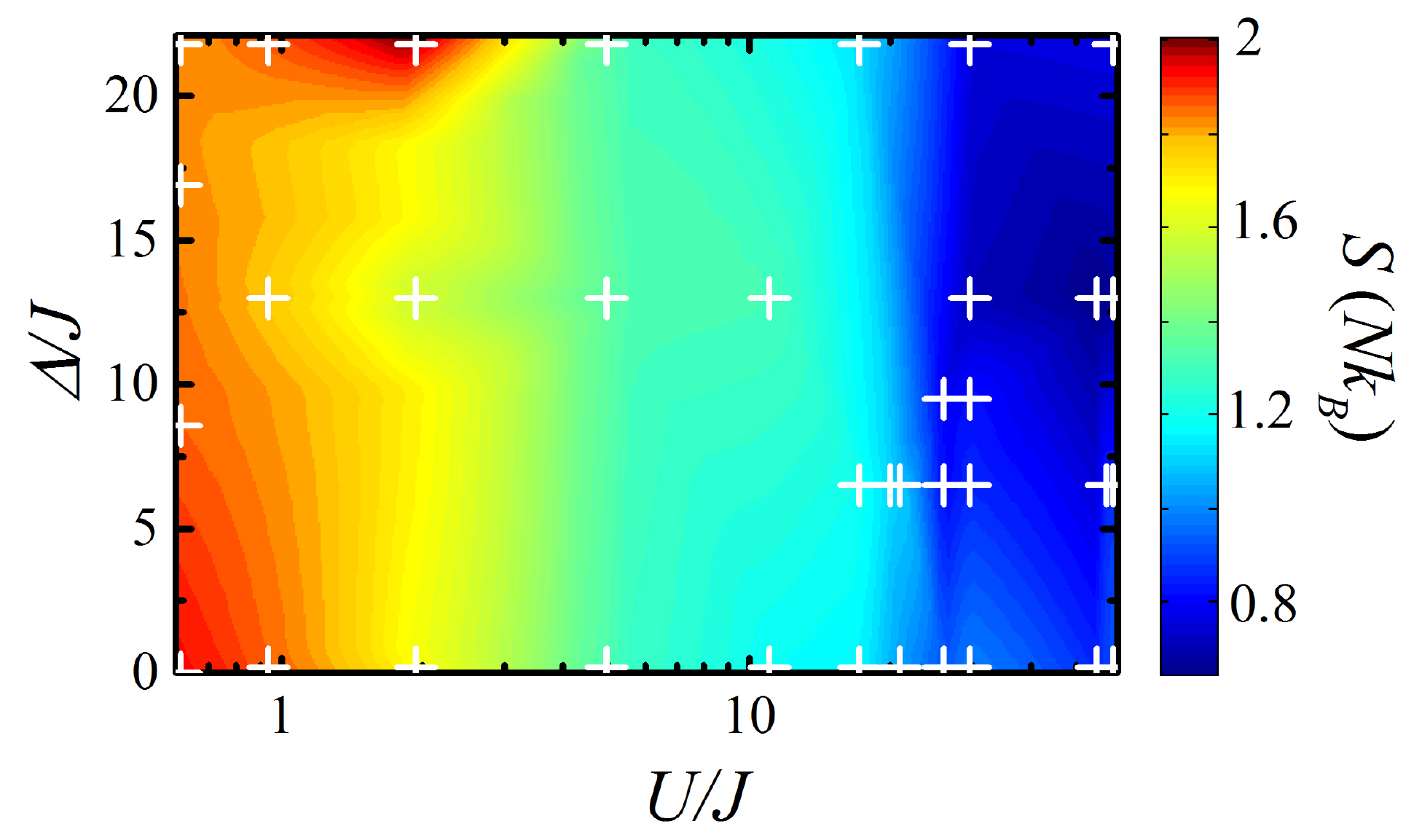}
\caption{$U$-$\Delta$ diagram for experimental estimates of the entropy per particle, $S/Nk_B$. The white crosses show the data points from which the 2D diagram was generated by interpolation. Data taken from supplemental material of Ref.~\cite{Derrico}.}
\label{figS9}
\end{figure}

In the absence of a straightforward thermometry procedure for the full diagram, we estimate the  experimental entropy, according to the procedure described in Sec.~\ref{sec:GammaVsS}, to provide an indication for the temperature changes with respect to the temperature estimates obtained for the clean case ($\Delta=0$). Fig.~\ref{figS9} shows the entropy $S$ of the system across the $U$-$\Delta$ diagram.
We observe that $S$ is quite independent of $\Delta$ and displays an overall increase towards small $U$,  which is presumably due to a reduced adiabaticity in the preparation of the 1D systems for weak interactions. This result is in agreement with the fact that the temperature estimated for the Mott-insulating regime is smaller than the one found for the superfluid in Sec.~\ref{sec:thermometry}. Moreover, the measurement suggests that an increase of disorder might likely not be accompanied by an increase of temperature.

\section{Conclusions}
\label{sec:conclusions}
The behavior of quantum matter  in the presence of disorder and interaction is a very complex subject, especially when one studies experimental systems which, beside being inhomogeneous due to the trap confinement, are necessarily at finite temperature. 
Starting from a recent study on the quantum phases observed in 1D bosonic disordered systems \cite{Derrico}, in this paper we provided a careful examination of the effects of finite temperature. To this purpose, two different DMRG schemes have been employed: (i) a direct simulation of the thermal density matrix in the form of a matrix product purification, and (ii) a less costly phenomenological method based on DMRG ground state data that are extended to finite temperatures by introducing an effective thermal correlation length. 
This analysis of our inhomogeneous system is  corroborated by  exact diagonalization studies for small-sized systems without trapping potential. While in the weak-interaction regime thermal effects can be rather strong, they are significantly less relevant in the strong-interaction one. There, the scaling of the correlation length with $T$ shows a weak dependence below a crossover temperature, indicating that the strongly-correlated quantum phases predicted by the $T=0$ theory can persist at the finite temperatures of our experiment. 

Furthermore, by using quasi-exact finite-$T$ DMRG simulations, we provided a temperature estimate for a superfluid in a lattice, the main source of uncertainty being the actual distribution of atoms among several quasi-1D systems in the experiment. Experimentally, a possible way to reduce this uncertainty is  to use a flat top beam shaper providing homogeneous trapped systems \cite{Veldkamp, Hoffnagle, Liang, Hadzibabic}.  The latter modification would for example also allow for a better discrimination of the features of the Bose glass  and the Mott insulator in the strong-interaction regime. 

In the insulating regimes, the Mott insulator and the Bose glass, experimental thermalization issues prevent precise temperature estimates.
A mixture with an atomic species in a selective potential \cite{Minardi} working as a thermal bath could be employed to guarantee thermalization of the species under investigation.
Another open question is whether the persistence of the insulating behavior for the disordered system with weak interactions could be related to the proposed many-body localization phenomenon \cite{Aleiner,Michal}. 

\subsection*{Acknowledgements}

This research was supported by EU FET-Proactive QUIC (H2020 grant No. 641122), Italian MIUR (grant No. RBFR12NLNA). 
GR acknowledges support from the French ANR program ANR-2011-BS04-012-01 QuDec.


\begin{thebibliography}{99}


\bibitem{Anderson}	P. W. Anderson, Phys. Rev. \textbf{109}, 1492 (1958).
\bibitem{Billy}  J. Billy, V. Josse, Z. Zuo, A. Bernard, B. Hambrecht, P. Lugan, D. Cl\'ement, L. Sanchez-Palencia, P. Bouyer, and A. Aspect, Nature (London) \textbf{453}, 891 (2008).
\bibitem{Roati08}	G. Roati, C. D'Errico, L. Fallani, M. Fattori, C. Fort, M. Zaccanti, G. Modugno, M. Modugno, and M. Inguscio, Nature \textbf{453}, 895 (2008).
\bibitem{Giamarchi88}	T. Giamarchi, H. J. Schulz, Phys. Rev. B \textbf{37}, 325 (1988).
\bibitem{Prokofev98} N. V. Prokof'ev, B. V. Svistunov, Phys. Rev. Lett. \textbf{80}, 4355 (1998).
\bibitem{Rapsch99}  S. Rapsch, U. Schollwöck and W. Zwerger, Europhys. Lett. \textbf{46}, 559 (1999).
\bibitem{Roth}	R. Roth, K. Burnett, Phys. Rev. A \textbf{68}, 023604 (2003).
\bibitem{Roscilde}	T. Roscilde, Phys. Rev. A \textbf{77}, 063605 (2008).
\bibitem{Deng} X. Deng, R. Citro, A. Minguzzi, E. Orignac, Phys. Rev. A \textbf{78}, 013625 (2008).
\bibitem{Roux}	G. Roux, T. Barthel, I. P. McCulloch, C. Kollath, U. Schollw\"ock, and T. Giamarchi, Phys. Rev. A \textbf{78}, 023628 (2008).
\bibitem{Modugno}  M. Modugno, New J. Phys.  \textbf{11},  033023 (2009).
\bibitem{Fallani}	L. Fallani, J. E. Lye, V. Guarrera, C. Fort, M. Inguscio, Phys. Rev. Lett. \textbf{98}, 130404 (2007).
\bibitem{Deissler}	B. Deissler, M. Zaccanti, G. Roati, C. D'Errico, M. Fattori, M. Modugno, G. Modugno and M. Inguscio, Nat. Phys. \textbf{6}, 354 (2010).
\bibitem{DeMarco} M. Pasienski, D. McKay, M. White, and B. DeMarco, Nat. Phys. \textbf{6}, 677 (2010).
\bibitem{Derrico} C. D'Errico, E. Lucioni, L. Tanzi, L. Gori, G. Roux, I. P. McCulloch, T. Giamarchi, M. Inguscio, and G. Modugno, Phys. Rev. Lett. \textbf{113}, 095301 (2014).
\bibitem{Michal} V. P. Michal, B. L. Altshuler, G. V. Shlyapnikov, Phys. Rev. Lett. \textbf{113}, 045304 (2014).
\bibitem{Gerbier}	F. Gerbier, Phys. Rev. Lett. \textbf{99}, 120405 (2007).
\bibitem{White1992} S.R. White, Phys. Rev. Lett. \textbf{69}, 2863 (1992).
\bibitem{White1993}  S.R. White, Phys. Rev. B \textbf{48}, 10345 (1993).
\bibitem{Schollwock}	U. Schollw\"ock, Rev. Mod. Phys. \textbf{77}, 259 (2005).
\bibitem{Verstraete} F. Verstraete, J.J. Garciıa-Ripoll, and J.I. Cirac, Phys. Rev. Lett. \textbf{93}, 207204 (2004).
\bibitem{Feiguin} A.E. Feiguin and S.R. White, Phys. Rev. B \textbf{72}, 220401(R) (2005).
\bibitem{Barthel} T. Barthel, U. Schollw\"{o}ck, and S.R. White, Phys. Rev. B \textbf{79}, 245101 (2009).
\bibitem{Binder} M. Binder and T. Barthel, Phys. Rev. B \textbf{92}, 125119 (2015).
\bibitem{Gerbier03} S. Richard, F. Gerbier, J. Thywissen, M. Hugbart, P. Bouyer, and A. Aspect, Phys. Rev. Lett. \textbf{91}, 010405 (2003).
\bibitem{Gerbier04} F. Gerbier, Condensats de Bose-Einstein dans un pi\`ege anisotrope, Ann. Phys. Fr. \textbf{29} 1 (2004).
\bibitem{Roati07}	G. Roati, M. Zaccanti, C. D'Errico, J. Catani, M. Modugno, A. Simoni, M. Inguscio, and G. Modugno, Phys. Rev. Lett. \textbf{99}, 010403 (2007).
\bibitem{Petrov}	D. S. Petrov, G. V. Shlyapnikov, J. T. M. Walraven, Phys. Rev. Lett. \textbf{85}, 3745 (2000).
\bibitem{Trotzky} S. Trotzky,	L. Pollet,	F. Gerbier,	U. Schnorrberger,	I. Bloch,	N. V. Prokof'ev,	B. Svistunov and M. Troyer, Nat. Phys. \textbf{6}, 998 (2010).
\bibitem{Giamarchibook}	T. Giamarchi, Quantum physics in one dimension (Clarendon, Oxford, 2004).
\bibitem{Aubry}	S. Aubry, G. Andr\'e, Ann. Israel Phys. Soc. \textbf{3}, 133 (1980).
\bibitem{Fisher}	M. P. A. Fisher, P. B. Weichman, G. Grinstein, D. S. Fisher, Phys. Rev. B \textbf{40}, 546 (1989).
\bibitem{Sirker}	J. Sirker, A. Klumper, Phys. Rev. B \textbf{66}, 245102 (2002).
\bibitem{Nessi}	 N. Nessi, A. Iucci, Phys. Rev. A \textbf{84}, 063614 (2011).
\bibitem{Aleiner}	I. L. Aleiner, B. L. Altshuler, G. V. Shlyapnikov, Nat. Phys. \textbf{6}, 900 (2010).
\bibitem{Catani}	J. Catani, G. Barontini, G. Lamporesi, F. Rabatti, G. Thalhammer, F. Minardi, S. Stringari, and M. Inguscio, Phys. Rev. Lett. \textbf{103}, 140401 (2009).
\bibitem{Delande} J. Zakrzewski and D. Delande, Phys. Rev. A \textbf{80}, 013602 (2009).
\bibitem{Bernier} J.-S. Bernier, G. Roux, and C. Kollath, Phys. Rev. Lett. \textbf{106}, 200601 (2011);
 J.-S. Bernier, D. Poletti, P. Barmettler, G. Roux, and C. Kollath, Phys. Rev. A \textbf{85}, 033641
(2012).
\bibitem{Veldkamp} W. B. Veldkamp.  Applied optics, 21(17):3209Ð3212 (1982).
\bibitem{Hoffnagle} J. A. Hoffnagle and C. M. Jefferson.  Applied optics, 39(30):5488Ð5499 (2000).
\bibitem{Liang} J. Liang, R. N. Kohn Jr, M. F. Becker, D. J. Heinzen, et al.  Applied optics, 48(10):1955Ð1962 (2009).
\bibitem{Hadzibabic} A. L. Gaunt, T. F. Schmidutz, I. Gotlibovych, R. P. Smith, and Z., Phys. Rev. Lett. \textbf{110}, 200406 (2013).
\bibitem{Minardi} J. Catani et al. Phys. Rev. A \textbf{85}, 023623 (2012).


\end{thebibliography}
\end{document}